\documentclass[11pt,a4paper]{article}
\usepackage{amsmath, amsfonts,amssymb,mathtools,nicefrac,nccmath,cases}
\usepackage{microtype,booktabs,multirow}
\usepackage[usenames,dvipsnames,table]{xcolor}
\usepackage{colortbl}

\def\ni{\noindent}
\def\foot{\footnote}
\def\fund{{\bf fund}}
\def\asym{{\bf asym}}
\def\sym{{\bf sym}}
\usepackage{tikz}
\usetikzlibrary{shapes,arrows}
\usetikzlibrary{calc,shapes.callouts,shapes.arrows,bending,intersections}
\usetikzlibrary{shapes.geometric}
\usetikzlibrary{arrows.meta,arrows}
\usetikzlibrary{decorations.pathmorphing}
\usetikzlibrary{decorations.markings}
\usetikzlibrary{shapes.multipart}
\usetikzlibrary{tikzmark}
\usepackage{arydshln}
\usepackage{caption}
\usepackage{subcaption}
\usepackage{amssymb}
\usepackage{amsmath}
\usepackage{physics}
\usepackage{amsmath}
\usepackage{tikz}
\usepackage{mathdots}
\usepackage{yhmath}
\usepackage{cancel}
\usepackage{color}
\usepackage{array}
\usepackage{multirow}
\usepackage{amssymb}
\usepackage{gensymb}
\usepackage{tabularx}
\usepackage{booktabs}
\usetikzlibrary{patterns}
\usetikzlibrary{shadows.blur}
\usetikzlibrary{shapes}
\usepackage{float}
\usetikzlibrary{fadings}
\usetikzlibrary{patterns}
\usetikzlibrary{shadows.blur}
\usetikzlibrary{shapes}
\usepackage[nosort]{cite}
\usepackage{colortbl}

\usepackage{cite}
\usepackage{enumerate}
\usepackage{ifpdf}
\usepackage{ae}
\usepackage[T1]{fontenc}
\usepackage[ansinew]{inputenc}
\usepackage{amsmath}
\usepackage{relsize}
\usepackage{amssymb}
\usepackage{comment}
\usepackage{graphicx}
\usepackage{color}
\definecolor{darkblue}{cmyk}{0.9,0.9,0,0}
\definecolor{darkgreen}{rgb}{0,0.55,0}
\usepackage{epsfig}
\usepackage{epstopdf}
\usepackage{mathtools}
\usepackage{graphicx}

\usepackage{varioref}
\usepackage{makeidx}
\usepackage{xcolor}
\usepackage{lipsum}

\usepackage[linktocpage=true]{hyperref}
\hypersetup{colorlinks=true,linkcolor=nicecolor,citecolor=nicecolor,urlcolor=nicecolor}
\definecolor{nicecolor}{rgb}{0.1, 0.3, 0.4}

\definecolor{blue}{rgb}{0.06, 0.3, 0.57}
\definecolor{Gray}{gray}{0.4}

\definecolor{nicecolor}{rgb}{0.1, 0.3, 0.4}
\definecolor{blue}{rgb}{0.06, 0.3, 0.57}
\definecolor{Gray}{gray}{0.4}
\colorlet{tableheadcolor}{gray!15} 
\colorlet{tablerowcolor}{gray!7} 

\def\hybrid{\topmargin -5pt    \oddsidemargin 0pt
	\headheight 0pt \headsep 0pt
	\textwidth 6.5in        
	\textheight 9in         
	\textwidth 6.25in       
	\textheight 9 in       
	\marginparwidth .875in
	\parskip 5pt plus 1pt 
	\jot = 1.5ex
}
\hybrid
\numberwithin{equation}{section}
\numberwithin{table}{section}
\usepackage{float}
\usepackage{tabularx}
\newcolumntype{D}{>{\centering\arraybackslash}X}
\newcolumntype{L}{>{$}l<{$}}
\newcolumntype{R}{>{$}r<{$}}
\newcolumntype{C}{>{$}c<{$}}
\usepackage{IEEEtrantools}
\usepackage{makecell}

\newcommand{\beq}{\begin{equation}}  \newcommand{\eeq}{\end{equation}}
\newcommand{\bal}{\begin{aligned}}   \newcommand{\eal}{\end{aligned}}
\newcommand{\bea}{\begin{eqnarray}}  \newcommand{\eea}{\end{eqnarray}}
\def\beqa{\begin{eqnarray}}
\def\eeqa{\end{eqnarray}}

\newcommand{\bmat}{\left(\begin{array}}
\newcommand{\emat}{\end{array}\right)}

\newcommand{\cN}{\mathcal{N}}

\newcommand{\cM}{\mathcal M}

\def\Im{\text{Im}}

\def\tb{{\bar\tau}}

\def\eps{\epsilon}

\def\M{\mathcal{M}}

\def\1{{\rm 1-loop}}

\def\Tr{{\rm Tr}}

\def\c{\cite}

\def\cM{\mathcal{M}}

\def\cN{\mathcal{N}}

\def\c{\cite}

\def\vs{\vskip .1 in}

\def\p{\partial}
\def\o{\over}
\def\g{\gamma}
\def\D{\Delta}
\def\rar{\rightarrow}

\def\eqr{\eqref}

\def\O{{\cal O}}
\def\ra{\rangle}
\def\la{\langle}
\def\ssec{\subsection}
\def\sssec{\subsubsection}
\def\sec{\section}
\def\i{\infty}

\newcommand{\e}[2] {\begin{equation} \label{#1} #2 \end{equation}}

\newcommand{\beqy} {\begin{eqnarray}}
\newcommand{\eeqy} {\end{eqnarray}}
\newcommand{\bsmat}{\begin{smallmatrix}}
\newcommand{\esmat}{\end{smallmatrix}}

\def\({\left(}
\def\){\right)}
\def\[{\left[}
\def\]{\right]}

\def\<{\langle}
\def\>{\rangle}

\def\a{\alpha}
\def\b{\beta}
\def\g{\gamma}

\def\D{\Delta}

\def\l{\lambda}

\def\t{\tau}

\def\vs{\vskip .1 in}

\usepackage{cancel}

\newcommand{\be}{\begin{equation}}
\newcommand{\ee}{\end{equation}}

\definecolor{Gray}{gray}{0.95}

\begin{document}

\baselineskip=14pt
\parskip 4pt

\vspace*{-1.5cm}
\begin{flushright}    
  {\small 
  
  }
\end{flushright}

\vspace{2cm}
\begin{center}        

  {\huge A CFT Distance Conjecture \\ 
   [.3cm]  }
 
\end{center}

\vspace{0.5cm}
\begin{center}        
{\large  Eric Perlmutter,$^{1}$ Leonardo Rastelli,$^{2}$ Cumrun Vafa$^{3}$ and Irene Valenzuela$^{3}$}
\end{center}

\vspace{0.15cm}
\begin{center}        
 \emph{$^{1}$Walter Burke Institute for Theoretical Physics, Caltech, Pasadena, CA 91125, USA }
 \\[.1cm]
  \emph{$^{2}$Yang Institute for Theoretical Physics, Stony Brook University, Stony Brook, NY 11793, USA}
   \\[.1cm]
 \emph{$^{3}$Jefferson Physical Laboratory, Harvard University, 
  Cambridge, MA 02138, USA }
 \\[.3cm]

\end{center}

\vspace{2cm}


\begin{abstract}
\noindent We formulate a series of conjectures relating the geometry of conformal manifolds to the spectrum of local operators in conformal field theories in $d>2$ spacetime dimensions. We focus on conformal manifolds with limiting points at infinite distance with respect to the Zamolodchikov metric. Our central conjecture is that all theories at infinite distance possess an emergent higher-spin symmetry, generated by an infinite tower of currents whose anomalous dimensions vanish exponentially in the distance. Stated geometrically, the diameter of a non-compact conformal manifold must diverge logarithmically in the higher-spin gap. In the holographic context our conjectures are related to the Distance Conjecture in the swampland program. Interpreted gravitationally, they imply that approaching infinite distance in moduli space at fixed AdS radius, a tower of higher-spin fields becomes massless at an exponential rate that is bounded from below in Planck units. We discuss further implications for conformal manifolds of superconformal field theories in three and four dimensions.

\end{abstract}

\thispagestyle{empty}
\clearpage

\setcounter{page}{1}


\newpage

\tableofcontents

\section{Introduction}

In the modern conformal bootstrap approach to conformal field theory (CFT), symmetries and consistency conditions are leveraged to quantitatively constrain the space of CFTs and properties thereof \c{Simmons-Duffin:2016gjk,Poland:2018epd}. A similar spirit animates the  approach to the landscape of theories of quantum gravity broadly known as the swampland program \c{Brennan:2017rbf,Palti:2019pca}. This work conjoins these perspectives to explore properties of conformal manifolds in CFT and their holographically dual moduli spaces in anti-de Sitter (AdS) space.\footnote{See \cite{Nakayama:2015hga,Harlow:2015lma,Benjamin:2016fhe,Montero:2016tif,Heidenreich:2016aqi,Montero:2017mdq,Harlow:2018tng,Bae:2018qym,Harlow:2018jwu,Lin:2019kpn,Montero:2018fns,Conlon:2018vov,Conlon:2020wmc,Ooguri:2020sua,Agarwal:2020pol} for some other works investigating the swampland conjectures using CFT techniques.} 

One of the important features of string theory is the existence of duality symmetries.  From the viewpoint of an effective theory this corresponds to the emergence of new descriptions of the physical system in terms of new light fields when we go to extreme limits of parameter spaces, e.g.~moduli spaces, which are captured by expectation values of scalar fields.  In such limits a dual description emerges leading to new weakly-coupled effective descriptions of the system.
In particular, effective field theories break down as we go from one extreme limit of scalar field vev's to another. This is a general feature of theories coupled to gravity, and so it has been proposed as a swampland criterion, known as the Distance Conjecture (DC) \cite{Ooguri:2006in} (or ``Duality Conjecture''), which is believed to encode consistency requirements of a quantum gravitational system. 

The DC in particular quantifies how quickly these light states emerge in extreme limits. 
Let $d(p,p_0)$ denote the shortest distance in moduli space between two expectation values $p$ and $p_0$ of the scalar fields, where distance is measured using the metric $h_{ij}$ defined by the kinetic term of the scalar fields in Einstein frame with action $S(\phi) = {1\over 2}\int d^D x \sqrt{g} h_{ij} \partial_\mu \phi^i \partial^\mu \phi^j$. Then the DC suggests that as $p$ goes to extreme points of the moduli space where $d(p,p_0)\gg1$, we get a tower of light states whose masses (in Planck units) vanish as 
\beq
m\sim {\rm e}^{-\alpha d(p,p_0)}\ ,
\eeq
where $\alpha$ is an $O(1)$ number \cite{Ooguri:2006in,Klaewer_2017}. This has been observed in many examples in string theory  \cite{Grimm:2018ohb,Lee:2018urn,Lee_2019,Gonzalo:2018guu,Grimm_2019,Corvilain:2018lgw,grimm2019infinite,Joshi:2019nzi,Marchesano_2019,Lee:2019tst,Lee:2019xtm,lee2019emergent,Font:2019cxq,Baume:2019sry,Blumenhagen:2017cxt,Blumenhagen_2018,Erkinger:2019umg,Cecotti:2020rjq,Gendler:2020dfp,Lanza:2020qmt,Klaewer:2020lfg}. Moreover, it is believed that there is a lower bound $\alpha\geq \alpha_D$ depending only on the dimension of spacetime, $D$.  This lower bound has been motivated in \cite{Bedroya:2019snp,Andriot:2020lea,Bedroya:2020rmd} from the Trans-Planckian Censorship Conjecture \cite{Bedroya:2019snp} and it is consistent with the bound obtained in four-dimensional $\cN=2$ theories arising from string theory compactifications \cite{Grimm:2018ohb,Gendler:2020dfp}. 
 
Most of the evidence for the DC comes from the study of the string landscape in flat space. The DC presumably also holds in AdS.  In particular, we expect that in a theory of AdS quantum gravity with moduli, extreme limits in moduli space are accompanied by towers of light states.\footnote{A generalization of the DC has also been considered in AdS space in \cite{Lust:2019zwm}, viewing the absolute value of the vacuum energy $|\Lambda|$ as such a parameter.  Using this generalized notion, it was proposed there that the nearly flat limit of pure AdS, where $\Lambda \rightarrow 0$, cannot exist, because in such a limit we always get a tower of light states with mass $m\sim |\Lambda|^{a}$.  In this paper we consider the usual application of DC where we vary the vev of scalar fields, for which $\Lambda$ in the Einstein frame does not vary.} The key to turning this into a more concrete proposal is to define what tower of operators is becoming light, and what exactly ``light'' means in the presence of an extra length scale, $L_{\rm AdS}$. Noting that the holographic dictionary maps the bulk moduli space to the space of exactly marginal couplings in CFT, this general intuition encourages a search for a precise conjecture about limits of conformal manifolds phrased purely in the language of CFT.  
 
The conformal bootstrap approach to constraining CFTs is most effective when the CFT universality class in question contains isolated fixed points, the canonical paradigm being that of the $d=3$ Ising CFT \c{ElShowk:2012ht,El-Showk:2014dwa,Kos:2016ysd,dsdi,Caron-Huot:2020ouj}. In contrast, a CFT with exactly marginal operators presents certain challenges. Such a CFT possesses a conformal manifold, call it $\cM$, parameterized by the exactly marginal couplings. That the CFT data depend on these free parameters {\it a priori} prevents the crossing symmetry-based approach from shrinking the allowed theory space down to a point: while the points on $\cM$ where CFT data are {\it extremized} as a function of the couplings lie on the boundary of allowed theory space, the functional dependence of CFT data across $\M$ is quite difficult to ascertain from the bootstrap. Such extremal points are interesting in their own right (e.g.~\c{Beem:2013qxa, Beem:2016wfs}), but it remains an outstanding problem for the conformal bootstrap to develop robust methods for solving CFTs with conformal manifolds. (See \c{Beem:2013qxa, Beem:2016wfs, Lin:2015wcg,Behan:2017mwi,Bashmakov:2017rko,Baggio:2017mas,Kaidi:2020ecu} for some initial approaches.) Of course, this is not just a bespoke bootstrap problem: the reasons to study conformal manifolds are manifold, with a long and rich history.

Narrowing the scope slightly, a reasonable and highly rewarding goal for the bootstrap approach to conformal manifolds would be to find explicit connections between metric data of $\M$ and the local operator spectral data of a CFT. A logical step forward is to focus on degeneration limits of $\M$. On general grounds, one might expect enhanced symmetries to play an important role at such points. A suggestive tip comes from the fact that free CFTs enjoy a higher-spin symmetry and that, in a sense to be recalled later, the converse is also true: all higher-spin CFTs are essentially free \c{Maldacena:2011jn}. 

Inspired by these bootstrap and swampland considerations in tandem (as well as some earlier work on the landscape of quantum field theory \c{Kontsevich:2000yf, KS, Acharya:2006zw, Douglas:2010ic}), we will present a series of conjectures about the properties of infinite distance limits of conformal manifolds as a function of the higher-spin spectrum. These conjectures are valid for generic $d>2$ CFTs, with arbitrary central charge. They are sustained by known data collected from the landscape of conformal manifolds of superconformal field theories (SCFTs), and yet they suggest a number of intriguing new properties of SCFTs, particularly those with reduced supersymmetry. Let us  summarize the core idea very briefly, leaving the precise statement of the conjectures for the next section. A natural notion of distance on $\M$, which we will use henceforth, is defined by the {\it Zamolodchikov metric,} the matrix of two-point functions of exactly marginal operators. The gravity intuition of the DC suggests that as one approaches infinite distance from the interior of $\M$, a tower of operators in the CFT saturates some bound. Our proposal is that this tower is always comprised of {\it higher-spin} operators of unbounded spin which saturate their respective unitarity bounds in the infinite distance limit, thus becoming conserved. In other words, the leading Regge trajectory becomes exactly linear with unit slope at infinite distance. Moreover, their anomalous dimensions vanish {\it exponentially} in the distance. This emergent symmetry is, we conjecture, a necessary and sufficient condition for $\M$ to have infinite diameter. The rest of this paper is devoted to formalizing these conjectures, and developing their consequences for AdS quantum gravity and the relation to the DC. 

The organization of this paper is as follows. Section \ref{s2} presents the CFT Distance Conjecture, which contains a few components. In Section \ref{susy} we gather some strong evidence for our conjecture by surveying  the landscape of SCFTs in various dimensions.  Section \ref{s3} phrases the conjecture in terms of quantum gravity in AdS and writes the exponential decay rate of anomalous dimensions in Planck units. In Subsection \ref{s31}, we observe a lower bound on the exponent and catalog its value in some SCFTs. Subsection \ref{s32} examines the interplay with the DC in the Swampland program, and Subsection \ref{s33} explores how the higher-spin towers are manifest in gravity duals of a few well-studied CFTs. 

\emph{Note added:} While finalizing this paper, we became aware of \cite{Baume:2020dqd}, which partially overlaps with our results.

\section{A CFT Distance Conjecture}
\label{s2}

We are interested in families of $d$-dimensional CFTs  with exactly marginal parameters. We assume the CFTs to be unitary and local, by which we mean in particular 
 that they have a local stress tensor operator.\footnote{In particular, the
two-point function coefficient $C_T$ of the canonically normalized stress tensor must be finite. This excludes from consideration e.g.~the infinite $N$ limit of  gauge theories. We will briefly comment on non-local CFTs at the end of this section.} 
The associated conformal manifold, ${\cal M}$, is endowed with a natural notion of distance between two points, namely, the geodesic distance with respect to the Zamolodchikov metric \cite{Zamolodchikov:1986gt},
\e{}{ {|x-y|}^{2d} \la \O_i(x)\O_j(y)\ra = g_{ij}(t^i)\, , }
where $\lbrace \O_i\rbrace$ are exactly marginal operators and $\lbrace t^i \rbrace$ are the associated local coordinates of ${\cal M}$. We will use these definitions of ``distance'' and ``metric'' henceforth. For reasons that will become clear, the case of $d=2$ must be distinguished from the case of $d>2$. We will focus on $d>2$ and make some brief comments about $d=2$ below.

In many examples, there are distinguished limiting points of ${\cal M}$ where a free subsector of the CFT decouples.\foot{By ``limiting point of ${\cal M}$'' we mean a CFT $\mathcal{T}_{t_*}$ defined by a convergent sequence of CFTs $\mathcal{T}_t$, with $t\in\M$, in the limit $t\rar t_*$.} The best understood class of examples are  $d=4$ SCFTs with ${\cal N}=2$ supersymmetry: these will serve as motivation and paradigm for our general conjectures. As we review below (Section \ref{N=2}), all examples of $d=4,\ {\cal N}=2$ conformal manifolds are parametrized locally by complexified gauge couplings and exhibit ``cusp" points where one or more gauge couplings are going to zero.  A typical
$d=4,\ {\cal N}=2$ conformal manifold exhibits several such cusps. This is often phrased as a statement of generalized S-duality \c{Argyres:2007cn, Gaiotto:2009we}: the same abstract SCFT is described by different weakly-coupled frames in different regions on ${\cal M}$. In the generic case, the limiting theory at the cusp becomes the direct sum of a free CFT and of an interacting SCFT. 

We can characterize these limiting points more abstractly, with no reference to a Lagrangian description nor supersymmetry, in terms of higher-spin (HS) symmetry enhancement. It is known that all free CFTs enjoy HS symmetry: that is, a free CFT contains at least one Regge trajectory comprised of an infinite tower of HS currents. These are local primary operators that furnish symmetric traceless representations of the Lorentz group of rank $J=2, 4, \dots$, whose conformal dimensions saturate the spin-$J$ unitarity bound,
\e{}{\Delta_J  \geq d-2+J~.}
The $J=2$ current is the stress tensor, which exists in any local CFT. Via \c{Maldacena:2011jn, Maldacena:2012sf} and various follow-ups \c{Stanev:2013qra, Boulanger:2013zza, Hartman:2015lfa, Li:2015itl, Alba:2015upa,Meltzer:2018tnm}, the converse is also known to be true. A more precise version of the main statement is as follows: in a unitary CFT with finite central charge and a unique stress tensor (i)~the presence of one current with $J >2$ implies the existence of an infinite tower of HS currents of unbounded spin, and (ii) the correlation functions of the HS currents are those of a free theory.\footnote{The works \c{Maldacena:2011jn, Maldacena:2012sf} proved these statements for symmetric traceless currents. In $d>3$ there exist operators in mixed-symmetry representations of the Lorentz group, and one needs a broader definition of HS currents.
E.g. in the physically relevant case of $d=4$ (the only dimension $d >3$ where SCFTs can admit a conformal manifold, see discussion in Section \ref{susy}), a HS current is a primary operator of Lorentz quantum numbers $(j_1, j_2)$ for  $SO(4) \cong SU(2)_1 \times SU(2)_2$ saturating the unitarity bound $\Delta = 2 + j_1 + j_2$, and with $j_1 + j_2 >2$. We will continue to phrase things in terms of symmetric traceless tensors, leaving implicit the more general situation in which $J$ is a multi-index.} This implies that, at the level of correlation functions, the CFT must be the direct sum of a free theory and a (possibly trivial) CFT$'$. We will refer to a limiting point of $\cM$ with HS symmetry enhancement as a {\it HS point}. An equivalent characterization of a HS point is as a CFT with an accumulation in the twist spectrum at $\mathfrak{t}=d-2$, where twist is defined as $\mathfrak{t}:= \D-J$.

In all examples of conformal manifolds we are aware of, for CFTs in any dimension $d > 2$, the geodesic distance between an HS point and any point on $\M$ is infinite. In other words, by the definition given in the previous sentence, HS points are ``at infinite distance.''\foot{As studied in \c{Kontsevich:2000yf,Roggenkamp:2003qp,Roggenkamp:2008jm, Douglas:2010ic} using the rigorous machinery of $d=2$ CFTs, different infinite distance limits may exist, not all of which need satisfy the axioms of quantum field theory. The claims in this work pertain only to limits which converge to bona fide CFTs. Note also that the limit of the sequence of points $t\in\M$ may not be strictly ``on" $\M$ (absent a compactification of the moduli space). For free-field limiting points, this manifests itself in more physical terms in two ways: first, the breakdown of conformal perturbation theory around free CFTs; and second, the number of independent stress tensors changes discontinuously in the strict decoupling limit. For these (somewhat formal) reasons, we sometimes refer to limiting points as being at infinite distance from $\M$.} A prelude to the CFT Distance Conjecture is that this is a general fact: 
\vs
{\bf Conjecture I:} {\it All HS points are at infinite distance.}
\vs
\ni 
We will prove Conjecture I for SCFTs in Section \ref{susy}. Our main conjecture, deeper but more adventurous, is the converse statement:
\vs
{\bf Conjecture II:} {\it All CFTs at infinite distance are HS points. }
\vs
\ni This can be phrased more quantitatively in terms of an upper bound on the {\it diameter} of ${\cal M}$, defined in the obvious way,
\begin{equation}
{\rm diam} \[ {\cal M}\] \coloneqq {\rm sup} \{ d (t_1, t_2) \, | \; \; t_1\, , t_2 \in {\cal M} \}\,.
\end{equation}
where $d(t_1,t_2)$ is the distance. Our claim is that the diameter is controlled by the anomalous dimensions of HS operators,
\e{}{\gamma_J  \coloneqq \Delta - (J +d-2)\,.}
 Let us introduce some further notation. Consider a family of CFTs ${\cal T}_t$ labelled by  $t \in {\cal M}$. The collection of Hilbert spaces ${\cal H}_t$ of  ${\cal T}_t$ defines a bundle. By the usual state/operator map, ${\cal H}_t$ is isomorphic to the space of local operators of ${\cal T}_t$. We can grade ${\cal H}_t$ by the spin quantum number,
 \begin{equation}
  {\cal H}_t =  \bigoplus_J    {\cal H}_t^{(J)}\, .
 \end{equation}
 We define $\gamma_J (t)$ to be the smallest anomalous dimension for operators of spin $J$ at point $t$ on $\M$,
 \begin{equation}\label{eps}
 \gamma_J (t)  \coloneqq {\rm min} \{ \g_\O \, | \;  {\cal O} \in    {\cal H}_t^{(J)}  \}\,.
 \end{equation}
Note that  $\gamma_2 (t) = 0$ for all $t$, because we assume the existence of a stress tensor.  The criterion for a limiting point to exhibit HS symmetry is  $\gamma_4 \to 0$, because a $J=4$ current is present in any HS algebra (whereas odd-spin currents need not be). Then a slightly weaker version of Conjecture II may be stated as follows:
\vs
{\bf Conjecture II$'$:} 
{\it  If  $\gamma_4 (t) > \epsilon$ for a fixed $\epsilon >0$, $\forall~t \in {\cal M}$, then ${\rm diam} \[ {\cal M}\]    < \infty$.}
\vs

 Next, let us consider how the HS points are approached. Suppose $\lim_{t \to t_0} \gamma_4 (t)= 0$ for some limiting point $t_0$. The behavior in several class of examples leads us to a yet more quantitative conjecture: the anomalous dimension near a HS point approaches zero {\it exponentially} in the distance. To motivate this from purely CFT considerations, consider the simple situation of a $d=4$, ${\cal N}=2$ gauge theory with a single exactly marginal complexified gauge coupling, $\t$. The metric is locally hyperbolic near the cusp \c{Papadodimas:2009eu},
\e{zmetric}{ds^2 \approx \b^2 {d\t d\tb \o (\Im \,\t)^2} \quad \text{as}\quad \Im \,\t\rar\i}
where $\b$ is a constant. The proper distance from an arbitrary point $\t'\in\M$ to the HS point is thus logarithmic,
\e{}{d(\t,\t') \approx \b \log \Im \,\t  \quad \text{as}\quad \Im \,\t\rar\i}
At $O(g^2)$, where $\text{Im}\,\t \sim g^{-2}$, all HS currents must acquire anomalous dimensions. This can be shown by explicit perturbative calculation in examples, but also follows abstractly from the following logic. At $O(g^2)$, if there is at least one local operator with nonzero anomalous dimension, the CFT is no longer free. The results of \c{Maldacena:2011jn, Maldacena:2012sf, Stanev:2013qra, Boulanger:2013zza, Hartman:2015lfa, Li:2015itl, Alba:2015upa,Meltzer:2018tnm} then imply that the HS symmetry must be broken. Furthermore, since the existence of a single HS current implies an infinite tower of HS currents, all HS currents must be broken at this order. Therefore, ignoring possible degeneracy at a given spin, the leading-order anomalous dimensions $\gamma_J$  of the HS currents are
\e{anomdim}{\gamma_J \sim f(J) \,g^2 \sim f(J)\exp\left(-{d (\t,\t') \o\b} \right)  \quad \text{as}\quad d(\t,\t')\rar\i}
for some analytic function $f(J)$. For example, in $d=4$, $\cN=4$ super-Yang-Mills (SYM) theory, the HS currents on the leading single-trace Regge trajectory obey \c{Anselmi:1998ms,Kotikov:2001sc,Dolan:2001tt}
\e{}{\gamma(J) \approx {g_{\rm YM}^2 N\o 2\pi^2 }H(J) + O(g_{ \rm YM}^4)\quad\text{where}\quad H(J) \coloneqq \sum_{n=1}^J {1\o n}\,.}
\ni If we make the assumption that \eqr{anomdim} is indeed the general asymptotic behavior near HS points -- equivalently, that the asymptotic metric approaching infinite distance is locally hyperbolic, \`a la \eqr{zmetric} -- then we can refine the CFT Distance Conjecture as follows.
We define ${\cal M}_\epsilon$ as the subset of the conformal manifold
where the anomalous dimension $\g_4$ is bounded below by $\epsilon > 0$,
\begin{equation}\label{Meps}
{\cal M}_\epsilon \equiv \{ t \in {\cal M} \, | \; \gamma_4 (t) \geq \epsilon \} \,.
\end{equation} 
Suppose $\M$ contains a HS point. We then conjecture that as $\epsilon \to 0$, the diameter diverges {\it logarithmically}:
\vs
{\bf Conjecture III:} {\it If \,${\rm inf} \{ \gamma_4 (t) \, | \; t \in {\cal M} \} =0$, then ${\rm diam} \[ {\cal M}_{\epsilon\rar 0}\]   \sim  \beta \,|\!\log  \epsilon |$~.}
\vs
\ni The constant $\beta$ is theory-dependent. Below we will derive $\b$ when $t$ is a complexified gauge coupling in $d=4$.

This conjecture admits a further refinement using conformal Regge theory. By definition, the set of operators ${\cal O} \in    {\cal H}_t^{(J)}$, with anomalous dimensions $\gamma_J(t)$ at point $t\in\cM$, comprises the leading Regge trajectory. The foregoing conjectures may be equally phrased in terms of the Regge slope of the leading trajectory; for example, Conjecture II is the statement that all CFTs at infinite distance have a {\it linear} leading trajectory with unit slope. In general, $\gamma_J(t)$ may be continued to an analytic function in $J$ \c{Caron-Huot:2017vep}. It was proven in \c{Costa:2017twz} that in any CFT, the leading trajectory is monotonic and convex:
\e{convex}{\gamma'_J>0\,,\quad \gamma''_J<0.}
Near HS points, the tower of HS currents with exponentially vanishing anomalous dimensions must obey \eqr{convex}. Following \eqr{Meps}, we define
\begin{equation}
{\cal M}_\epsilon^{(J)} \coloneqq \{ t \in {\cal M} \, | \; \gamma_J (t) \geq \epsilon_J \} \,.
\end{equation} 
From the discussion earlier, $\epsilon_J \propto \eps$ as $\eps \rar 0$, and ${\rm diam} \[ {\cal M}_{\epsilon \rar 0}^{(J)}\]$ diverges logarithmically. Together with Conjecture III, \eqr{convex} implies that as $\epsilon \to 0$, there is a monotonicity and convexity condition on the subsets of the conformal manifold graded by spin: if ${\rm inf} \{ \gamma_4 (t) \, | \; t \in {\cal M} \} =0$, then ${\rm diam} \[ {\cal M}_{\epsilon \rar 0}^{(J)}\]$ should be monotonic and convex as a function of $J$.

Our Distance Conjectures assume that the CFT is local, i.e.~that it has  a local stress tensor. Our whole discussion  also relies on the standard geometric picture of the conformal manifold: the exactly-marginal local operators of ${\cal T}_t$, where
$t \in  {\cal M}$,
define the tangent space of     ${\cal M}$ at  point $t$. 
 This can also be understood as a statement of locality -- correlation functions at nearby points of ${\cal M}$ are related by standard conformal perturbation theory, i.e.~by integrating insertions of the local exactly marginal deformations. Conformal manifolds for local CFTs are very constrained -- e.g., as we shall review below, the only known examples 
in $d >2$ occur in supersymmetric theories.
On the other hand, it is easy to construct continuous families of CFTs if one gives up locality, see e.g.~\cite{Behan:2017emf, DiPietro:2019hqe}. 
The simplest (indeed trivial) example is just  ``generalized free-field theory" (GFFT),  the mean field theory
generated by a ``single-trace'' primary operator, e.g.~a scalar primary ${\cal O}_\Delta$ of conformal dimension $\Delta$, and  its arbitrary ``multi-trace'' composites, assuming exact large $N$ factorization.  
The dimension $\Delta$ is a free continuous parameter;
as $\Delta \to \frac{d}{2} -1$ we approach the usual local CFT of a free scalar, which enjoys HS symmetry. For $\Delta > \frac{d}{2} -1$, GFFT is unitary but non-local: there is no local stress tensor operator nor a local exactly marginal operator that could be identified as the tangent vector 
to the conformal manifold. More elaborate and physically interesting examples of continuous families of non-local, non-supersymmetric  CFTs are given in  \cite{Behan:2017emf, DiPietro:2019hqe}. They all lie outside the framework of our conjectures,
indeed it is not even clear how to define a natural notion of distance, or of a conformal manifold, in the absence of local exactly marginal deformations. Locality appears to be essential for  the CFT Distance Conjecture. When reinterpreted holographically, this statement is in nice agreement with the expectation that swampland constraints require the presence of dynamical gravity. 

In $d=3$, there are many fascinating sequences of Chern-Simons-matter CFTs admitting large $N$ 't Hooft limits, e.g.~\c{Schwarz:2004yj, Gaiotto:2007qi, Aharony:2008ug, Aharony:2008gk, Chang:2010sg, Giombi:2011kc,Aharony:2011jz,Chang:2012kt}. The prototypical 't Hooft coupling takes the  form $\l=N/k$, where $k$ is the Chern-Simons level. Since $k$ is quantized, $\l$ is not a  continuous parameter away from large $N$, and hence does not parameterize a conformal manifold. Certain vector models of this type also include a scalar $\phi^6$ coupling which, while exactly marginal at infinite $N$ \c{Aharony:2011jz}, ceases to be so at finite $N$. 

Finally, let us make some comments on the $d=2$ case. The discussion needs to be modified, for several reasons. First, from the perspective of the global conformal symmetry in $d=2$, local CFTs {\it always} have HS currents, constructed from composites of the holomorphic stress tensor. To formulate a criterion based on HS symmetry, one might instead require that the relevant HS currents be Virasoro primaries, but this does not appear to be useful either: composites of the Virasoro primary HS currents are themselves Virasoro primary, and they, too, are infinite in number. Going one step further, one might use the number of strong generators of the HS algebra (i.e.~those currents which cannot be written as composites) as a HS criterion; however, in $d=2$ there {\it do} exist HS algebras that are finitely generated. Finally, and most relevant for us, we know of no reason to associate HS points with infinite distance, e.g CFTs with HS chiral algebras need not be free.\foot{The $W_N$ minimal models are familiar examples.} Instead, it appears that the best $d=2$ analog of the CFT Distance Conjecture makes reference to the {\it scalar gap}, i.e.~$\gamma_0$. Limiting points with $\gamma_0 \to 0$ are at infinite distance (for example, large volume limits of Calabi-Yau sigma models). Kontsevich and Soibelman \c{Kontsevich:2000yf, KS} and Acharya and Douglas \cite{Acharya:2006zw} have conjectured that the converse is true: the only way to go to infinite distance is if $\gamma_0 \to 0$. 

Nevertheless, both the $d=2$ conjecture in \c{Kontsevich:2000yf, KS,Acharya:2006zw} and our $d>2$ conjectures involve infinite towers of operators approaching unitarity bounds at infinite distance. Therefore, they can be recast as the following universal but weaker statement applicable to $d\geq 2$: at infinite distance, there is an infinite tower of operators whose anomalous dimensions vanish. As we will note in Section \ref{s32}, in the holographic context this maps nicely to the DC in the Swampland program, in which there are infinite towers of massless fields with no universal constraint on their spin. 

\sec{Conformal manifolds for SCFTs}

\label{susy}

In dimension greater than two, all known examples of conformal manifolds occur in superconformal field theories (SCFTs) with at least four real Poincar\'e supercharges.\footnote{Even in $d=2$, non-supersymmetric CFTs with conformal manifolds are sparse. The only known examples are toroidal orbifolds.}
It is currently not understood whether this is  a lamppost effect or a deep fact about quantum field theory.
One may hope to show by  abstract bootstrap methods that the existence of a conformal manifold  implies supersymmetry, e.g.~by leveraging the intricate constraints on  the spectrum and OPE coefficients  imposed by the existence of exactly marginal operators~\cite{Bashmakov:2017rko, Behan:2017mwi},  
or  perhaps to find a non-supersymmetric example. With present technology, this appears to be a difficult problem. In this section, our more modest aim will be to support our conjectures by surveying the landscape of SCFTs in various dimensions. In doing so we will prove Conjecture I for SCFTs.

  Considerations of superconformal representation theory \cite{Cordova:2016emh, Cordova:2016xhm,Louis:2015mka}
restrict the number of preserved supercharges on the conformal manifold. Superconformal field theories 
 in $d> 2$ that can admit conformal manifolds belong to the following list:
\begin{itemize}
\item  In $d=3$,  SCFTs with ${\cal N}  = 2$  or (in principle) ${\cal N}=1$ supersymmetry.\footnote{\label{3dN=1foonote}
There are no known examples of conformal manifolds in $d=3$, $\cN =1$ SCFTs. This amount of supersymmetry is too little to establish the existence 
of exactly marginal operators. They are in principle allowed by superconformal representation theory, but they are top components of long supermultiplets; as such, one does not expect them to remain marginal beyond leading order in conformal perturbation theory, barring some deeper dynamical reason. In this regard, $d=3$, ${\cal N}=1$ SCFTs appear  to be on a similar footing as  non-supersymmetric CFTs. }
 If ${\cal N}>2$, there are no marginal operators preserving the additional supercharges, though supersymmetry enhancement is allowed at isolated points of the conformal manifold.  
\item  In $d=4$,  SCFTs with ${\cal N}=1$, ${\cal N}=2$ or ${\cal N}=4$ supersymmetry. A genuine ${\cal N}=3$ SCFT is necessarily an  isolated fixed point. There are many examples of ${\cal N}=1$ conformal manifolds where supersymmetry enhances to ${\cal N}=2$ or ${\cal N}=4$ on submanifolds. 
\end{itemize}
Superconformal field theories in $d = 5$ and $d=6$ are always isolated fixed points, and of course there are no SCFTs in $d>6$~\cite{Nahm:1977tg}.

We will restrict attention to SCFTs with at least four supercharges, i.e.~$d=3$, $\cN=2$ theories and $d=4$, $\cN=1, 2, 4$ theories. In all these cases, supersymmetry implies that the conformal manifold is complex, and that the Zamolodchikov metric
is K\"ahler (see e.g.~\cite{Asnin:2009xx}).

\ssec{$d=4$, $\cN =2$}
\label{N=2}

The best understood  class of examples of conformal manifolds in $d > 2$ occur in four-dimensional SCFTs with $\cN = 2$ supersymmetry.  

Some structural properties  can be established   from general principles.  It is easy to argue from  representation theory of the $\frak{su}(2, 2 |2)$ superconformal algebra (see e.g.~\cite{Dolan:2002zh, Cordova:2016emh, Cordova:2016xhm}) that $\cN =2$ exactly marginal deformations\footnote{We focus here on deformations that preserve the full $\cN =2$ supersymmetry. The $\cN =1$ conformal manifold is often larger, and will be discussed below.}
must take the form
\be
\delta S  =  \frac{1}{4 \pi^2}  \int d^4 x  \left(  \delta \tau^i \,  {\cal Q}^4  \phi_i +  \delta \bar \tau^{\bar \imath}  \,  {\overline{\cal Q}}^4 \overline{\phi}_{\bar \imath} \right) \, ,  
\ee
where $\phi_i$ and $\overline\phi_i$ are $\cN = 2$ chiral and antichiral primaries of $U(1)_R$ charge $R= \pm 4$ and scaling dimension\footnote{Our   conventions for the supersymmetry algebra follow \cite{Baggio:2014ioa}: ${\cal I}=1, 2$  are $SU(2)_R$ indices, 
$\alpha = \pm$ and $\dot \alpha = \dot \pm$ spinor indices;  the normalization of $R$ is fixed by $R[{\cal Q}] = -1$ and $R[\overline{\cal Q}] = +1$. We  also assume that  $\overline {\phi}_{\bar \imath}$ is the hermitian conjugate of   ${\phi}_{\bar \imath}$, and
  $\delta \bar \tau^{\bar \imath}$
the complex conjugate of $\delta \tau^i$, as required by unitarity.}
 $\Delta =|R|/2 = 2$,
\be
\[{\cal Q}^{\cal I}_\alpha, \overline \phi_i \]  =    \[\overline{\cal Q}^{\cal I}_{ \dot \alpha},   {\phi}_i \]  =  0  \, ,\quad   R \[  \phi_i  \] = - R\[  \overline\phi_i  \]  = 4\, .
\ee
The Zamolodchikov metric is given by
\be
g_{i \bar \jmath}  =  |x-y|^8  \langle  {\cal Q}^4  \phi_i  (x)   {\overline{\cal Q}}^4\,  \overline{\phi}_{\bar \jmath}  (y)  \rangle = \partial_i \partial_{\bar \jmath} {\cal K} \, ,
\ee
where ${\cal K}$ is the K\"ahler potential. A remarkable property of  $d=4$, $\cN =2$ SCFTs (for which however we will have little use in our analysis) is that one can identify the K\"ahler potential with the logarithm of the $S^4$ partition function,
${\cal K} = 192 \log Z_{S^4}$ \cite{Gerchkovitz:2014gta}. What's more, the conformal manifold ${\cal M}$   is K\"ahler-Hodge~\cite{Gomis:2015yaa} and its K\"ahler class can be determined from anomaly considerations \cite{Tachikawa:2017aux, Seiberg:2018ntt}. Supercharges 
are  associated to a holomorphic line bundle $\mathcal{L}$ over ${\cal M}$, whose curvature is proportional to the K\"ahler form of the Zamolodchikov metric~\cite{Papadodimas:2009eu}.\footnote{$\cN =2$ chiral operators can also be viewed 
as sections of a holomorphic line bundle on ${\cal M}$ with an interesting geometric structure~\cite{Papadodimas:2009eu}.}

In all {\it known} examples of $\cN= 2$ conformal manifolds, the exactly marginal parameters 
arise  as holomorphic gauge couplings, one for each simple factor of the total gauge group, $G_{\rm tot} = G_1 \times G_2 \dots G_n$.
 To wit,  even in non-Lagrangian SCFTs, the only known mechanism to generate $\cN =2$ exactly marginal couplings is by gauging  a subgroup  of the global symmetry group of a collection of isolated SCFTs, in such a way that the  beta function for each gauge coupling vanishes.\footnote{Non-renormalization theorems
  ensure that it is sufficient to check the vanishing of the one-loop beta function. To have a chance at a zero beta function, each  $G_i$ must be non-abelian.} All Lagrangian examples and all theories of class ${\cal S}$ \cite{Gaiotto:2009we} are of this type, with the isolated SCFTs being  just a set of free hypermultiplets in the Lagrangian case,
  and more general ``matter'' SCFTs  in the class ${\cal S}$ case.   It has been conjectured \cite{Beem:2014zpa} that this ``decomposability" in elementary isolated building blocks (glued by gauging)
is  a general fact about the
 landscape of $\cN = 2$ SCFTs, but a proof is lacking.

 At a generic point of ${\cal M}$ the metric is generally hard to compute,\footnote{One can sometimes leverage the relation of the K\"ahler potential with the $S^4$ partition function, 
 which in Lagrangian examples can in principle be evaluated using the techniques of supersymmetric localization \cite{Pestun:2007rz, Baggio:2014ioa, Baggio:2015vxa, Gerchkovitz:2016gxx, Grassi:2019txd}.}
but fortunately  we are  mostly concerned with its behavior  near  each of the weakly-coupled ``cusps" where a coupling $\tau \to i \infty$, as these are HS limiting points where vector multiplets are becoming free.
 Let us focus (just for notational simplicity) on a single simple factor $G$ of the total gauge group, and on its  coupling 
$\tau$.
 Near the cusp, we identify the  chiral primary $\phi$ as $\Tr \,\varphi^2$, where $\varphi$ is the complex scalar of the $\cN =2$ vector multiplet in the adjoint representation of $G$, 
  and the marginal operator with ${\cal O}_\tau \sim \Tr (F^2 + \dots) $, where $F_{\alpha \beta}$ is the self-dual field strength and the dots denote higher-order terms in the Yang-Mills coupling.\footnote{Similarly, the (CPT conjugate) antichiral primary is given by
 $\overline\phi  = \Tr \, \bar \varphi^2$ and the corresponding marginal operator by ${\cal O}_{\bar \tau} \sim \Tr ({\overline F}^2+ \dots)$, where ${\overline F}_{\dot \alpha \dot \beta}$ is the anti-selfdual field strength.} In the vicinity of the cusp, $\tau$ is then interpreted
 as the usual holomorphic gauge coupling,
 \be \label{hologauge}
 \tau =  \frac{4 \pi i}{g^2} + \frac{\theta}{2 \pi} \, .
 \ee
 As we approach the cusp from the interior of ${\cal M}$, the leading-order contribution to two-point function $\langle {\cal O}_\tau   {\cal O}_{\bar \tau} \rangle$  as $\tau \to i \infty$ is simply found  by performing tree-level Wick contractions, and thus
 $g_{\tau \bar \tau} \sim   g^4 ({\rm dim}\, G )$. 
In the normalization conventions  of  \cite{Baggio:2014ioa}, one finds
\begin{equation} \label{gaugecusp}
ds^2 \approx  
\beta^2 {d\t d\tb \o (\Im \,\t)^2} \quad \text{as}\; \Im \,\t\rar \infty \, , \quad \beta^2 = 24 \,  {\rm dim}\, G.
\end{equation}
Note that the local hyperbolic behavior of the metric  follows from a  perturbative argument. We also see that near the free vector point
 the metric depends only on the gauge group $G$ that is becoming free -- not on the rest of the SCFT -- and is proportional to its dimension.

Having set the stage, let us now examine our conjectures. Conjecture I follows at once. Suppose that ${\cal M}$ admits a HS  point  at finite distance.
Higher-spin symmetry implies the existence of a free decoupled sector, which in an $\cN =2$ SCFT means a collection of free hypermultiplets and/or free vector multiplets.
We can argue against the possibility that {\it  only} hypermultiplets become free using representation theory: the free hyper SCFT has no candidate composite operator with the
 quantum numbers\footnote{The conclusion still holds if one allows for the requisite composite operator to be the product of an operator from the free hyper SCFT and an operator from the decoupled SCFT' -- there is no candidate with the correct quantum numbers. }
of the chiral primary $\phi$. Therefore, the HS point must include free vectors. But as the analysis leading to \eqr{gaugecusp} shows, such a point lies an infinite distance away, with a logarithmic divergence of the diameter as a function of the HS anomalous dimensions.

Now we consider Conjecture II. This would follow from the above analysis if we could show that the {\it only} way to go to infinite distance is by approaching a limiting point with free vectors.  (Conjecture III would then be an easy consequence of the local hyperbolic behavior of the metric). This is harder to establish, but perfectly plausible. 
Starting from a generic point of ${\cal M}$, we can move to a nearby point by  conformal perturbation theory.\footnote{While not strictly needed for our considerations, a piece of lore which would be nice to establish more firmly is that conformal perturbation theory at an interior point of ${\cal M}$ has  finite radius of convergence, see e.g.~\cite{Kastor:1988ef}.} For the distance to diverge, we need  conformal perturbation theory to break down. This is indeed what happens at the free vector points:  Feynman perturbation theory 
applies,\footnote{Even in cases where the ``matter'' SCFT that is being gauged is  strongly coupled, we can adopt a hybrid formalism where the gauge field degrees of freedom, which have a standard Lagrangian description,  are coupled to the ``abstract" conserved current $J_\mu$ of the strongly-coupled SCFT, as e.g.~in \cite{Argyres:2007cn, Meade:2008wd}.} 
 but it is outside the standard framework of conformal perturbation theory, in that the perturbing operator (the supersymmetrization of $A_\mu J^\mu$) is not part of the  physical spectrum.
Our Conjecture II  would follow if one could show this is the {\it only} way for conformal perturbation theory to break down. If one were to find examples of $\cN =2$ SCFTs with exactly marginal deformations that could not be interpreted as holomorphic gauge couplings -- that is, if the SCFT were not decomposable -- then we are predicting that such exotic deformations would correspond to compact directions in ${\cal M}$. It is an interesting question whether the circle of ideas developed in this paper, possibly combined with the geometric constraints studied in \cite{Papadodimas:2009eu, Gomis:2015yaa, Tachikawa:2017aux, Seiberg:2018ntt}, may lead to a general proof of the decomposability conjecture for $\cN =2$ SCFTs.

 We can treat $d=4$, $\cN =4$ SCFTs as special cases of $\cN =2$ theories. The  conformal manifold of an irreducible $\cN = 4$ theory has complex dimension one, because the exactly marginal deformation is the top component of the stress-tensor multiplet.
 One can argue abstractly that ${\cal M}$ is locally a homogeneous hyperbolic manifold~\cite{Papadodimas:2009eu}, but the global structure is {\it a priori} unknown, e.g.~we do not know of an abstract argument that it should be non-compact.
 All known examples of $\cN =4$ SCFTs are  of course SYM theories, specified by a choice of simple gauge group $G$ and complexified gauge coupling $\tau$. The global structure is determined by the S-duality group, e.g.~for $G$ simply laced
 ${\cal M}$ is the fundamental domain of $SL(2, \mathbb{Z})$, endowed with the exact hyperbolic metric (\ref{gaugecusp}).

\ssec{$d=4$, $\cN=1$}
\label{N=1}

Conformal manifolds are ubiquitous in $\cN =1$ SCFTs. Their study was systematized in the classic paper of Leigh and Strassler \cite{Leigh:1995ep}. A simpler recipe to identify the space of exactly marginal deformations of a given $\cN =1$ SCFT
(i.e.,  locally on ${\cal M})$ was found by Green et al.~\cite{Green:2010da}. For a sample of the recent literature see e.g.~\cite{Razamat:2019vfd, Razamat:2020gcc, Razamat:2020pra, Komargodski:2020ved}.
Superconformal representation theory informs us that an $\cN =1$ preserving marginal operator must be the descendant of a chiral scalar operator\footnote{Here we follow the usual conventions for $\cN =1$ supersymmetry, with $r({\cal Q}) = -1$,  $r({\overline\cal Q}) = 1$. Chiral operators have dimension $\Delta = \frac{3}{2} r$.}
 with $U(1)_r$ charge $r = 2$ and scaling dimension $\Delta=3$,
\be
\delta S  =    \int d^4 x  \left(  \delta t^i \,  {\cal Q}^2  \chi_i  + {\rm h.c.} \right) \, , \quad \[\overline{\cal Q}_{ \dot \alpha},   {\chi}_i \]  =  0  \,, \;  r \[  \chi_i  \] = 
 2\, .
 \ee
Starting from a reference SCFT, which we call ${\cal T}_P$ (we think of $P$ as a point of ${\cal M}$), one  can enumerate all chiral scalar operators $\{ \chi_i \}$ with $r = 2$, but in general  only a subset of them is exactly marginal. The rest  are marginally irrelevant: they acquire a positive anomalous dimension away from $P$, becoming part of a long multiplet by recombining with a current multiplet. A careful analysis~\cite{Green:2010da} shows that 
 the conformal manifold in a neighborhood of  $P$ is given by the K\"ahler quotient
\be \label{recipe}
{\cal M} = \{  t^i \} / G_{\rm global}^{\mathbb C}\, ,
\ee
where $G_{\rm global}$ is the continuous global symmetry group of  ${\cal T}_P$, and  $ G_{\rm global}^{\mathbb C}$ its complexification. Properly interpreted, the recipe works both when  ${\cal T}_P$ does not contain free non-abelian\footnote{Abelian gauge fields cannot occur as their beta function would be positive, unless they are completely decoupled  from the rest of theory.}  gauge fields
 and when it does,
but the two situations are physically quite different:
\begin{itemize}
\item[(i):]
${\cal T}_P$ does {\it not} contain free  non-abelian gauge fields. The candidate marginal operators $\{ \chi_i \}$  are interpreted as ``matter" superpotential deformations.\footnote{Here and below we use ``matter''  to indicate a collection of free chiral multiplets, a strongly-coupled SCFT, or any combination  thereof.}
 We expect conformal perturbation theory to hold. In particular, a small deformation in an exactly marginal direction (\ref{recipe}) should cost a small  distance in the Zamolodchikov metric, i.e.~$P$ is a  regular  interior point of ${\cal M}$.
\item[(ii):] ${\cal T}_P$ contains free non-abelian gauge fields. It then
consists of a free gauge multiplet $W_\alpha$ in the adjoint representation of  $G_{\rm gauge}$, and of a decoupled ``matter'' sector with 
global symmetry $G_{\rm total}$.  One gauges a subgroup $G_{\rm gauge} \subset G_{\rm total}$. The recipe (\ref{recipe}) applies, if one  includes the complexified gauge coupling $\tau$ in the list of $\{ t^i \}$, and interprets
$G_{\rm global}$  as the centralizer\footnote{$G_{\rm global}$ is the global symmetry group at zero gauge coupling. The coupling to gauge fields may render a subgroup of $G_{\rm global}$  anomalous, so that it acts non-trivially on $\tau$.}
 of $G_{\rm gauge}$ in $G_{\rm total}$. However, as we have already remarked in the context of $\cN =2$ SCFTs, the gauging procedure falls outside the framework of conformal perturbation theory, and the deformation by $\Tr\, W_\alpha W^\alpha$ cannot be considered ``small''. We can repeat almost verbatim the analysis that led to (\ref{gaugecusp}),
and find exactly the same answer (in the usual normalization (\ref{hologauge}) of the holomorphic gauge coupling),
\begin{equation} \label{gaugecuspN=1}
ds^2 \approx 
\beta^2 {d\t d\tb \o (\Im \,\t)^2} \quad \text{as}\; \Im \,\t\rar \infty \, , \quad \beta^2 = 24 \,  {\rm dim}\, G.
\end{equation}
The Zamolodchikov metric has local hyperbolic behavior near the  limiting point with free gauge fields, which is then an infinite distance away.
\end{itemize}

Let us now examine our distance conjectures in the light of this analysis.   Mimicking the $\cN=2$ argument of the previous subsection, Conjecture I follows from the fact that no HS point at finite distance on ${\cal M}$ can have a free sector with {\it just} free chiral superfields. Suppose that this unwanted scenario were realized, i.e.~that as we approach the interior point $P$ we have HS symmetry enhancement,  with ${\cal T}_P$ the direct sum of a collection of free chiral superfields and of a decoupled theory ${\cal T}'$.  There is  a $U(1)$ global symmetry, $U(1) \subset G_{\rm global}$, which rotates all chiral superfields by a common phase. 
It then follows  from (\ref{recipe}) that there cannot be any exactly marginal operator $\chi_i$ that receives a contribution for the free-chiral sector,\footnote{The candidate deformation may be the product of free chiral superfields, or the  product of a free-chiral piece times a chiral operator of ${\cal T}'$.} as there is no way to cancel the $U(1)$ charge. All exactly marginal deformations must be part of ${\cal T}'$,
but then they do not move us away from the HS symmetry locus, against the original assumption.\footnote{Note that this argument doesn't work if  the free sector contains vector superfields as well: the holomorphic gauge coupling $\tau$ is {\it charged} under the $U(1)$
(i.e.~the $U(1)$ is rendered anomalous by the coupling to the gauge fields) and  it may be possible to find invariant combinations of the couplings.} Then Conjecture I follows from the argument leading to \eqr{gaugecuspN=1}, which shows that a HS limiting point containing free gauge fields lies at infinite distance.

Conjecture II (the only way to go to infinite distance is by approaching a HS point)
is harder, and we can only repeat what we said in the $\cN =2$ case: it would follow if we could show that 
the only mechanism for conformal perturbation theory to break down is by approaching a free vector point. The local hyperbolic behavior (\ref{gaugecuspN=1}) would then easily imply Conjecture III as well.

As our discussion has been quite abstract so far, let us illustrate it with a few concrete examples of $\cN=1$ conformal manifolds. 

Let us start with a classic example of type (ii), the $\cN =1$ conformal manifold of ${\cal N}=4$ SYM \cite{Leigh:1995ep} (see \cite{Aharony:2002hx} for a nice discussion). 
Taking the free-field point as reference point, ${\cal T}_P$ consists of the non-abelian gauge multiplet\footnote{We assume that the gauge group is a simple non-abelian  group different from $SU(2)$.} and of three adjoint chiral superfields $\Phi^i$, $i=1, 2, 3$. The global symmetry group
is $G_{\rm global} = U(3)$.  There is one complexified gauge coupling $\tau$ and 11 cubic superpotential couplings, which split as the ${\bf 1} +{\bf 10}$  irreps of SU(3),
\be
W =\frac{ \tau }{8 \pi i} \,\Tr \, W_\alpha W^\alpha + \frac{\tilde h}{6} \epsilon_{ijk}  \Tr \, \[\Phi^i \Phi^j \Phi^k  \] + \frac{1}{3} h_{ijk}   \Tr \, \[\Phi^i \Phi^j \Phi^k  \]  \, , 
\ee
with $h_{ijk}$ a fully symmetric tensor. A quick counting (which is easily confirmed by a careful construction of the K\"ahler quotient) gives $12- {\rm dim} \, U(3) = 12- 9 = 3$ exactly marginal couplings. The choice
 $\tilde h = g$ (where $g$ is the Yang-Mills coupling)  and
  $h_{ijk} = 0$  preserves the  full $\cN = 4$ supersymmetry. By a suitable $SU(3)$ rotation, one can  parametrize the general exactly marginal deformation by $\tau$ and  two superpotential couplings\footnote{The precise statement is that ${\cal M}$ is found
  by imposing one complex constraint for  four complex couplings, $f(\tau, \tilde h, h_1, h_2) =0$. What's more, at generic point on the 
 $\cN =4$ fixed line (parametrized by $\tau$), the two deformations (\ref{h12}) are exactly marginal  to leading order in $h_1$ and $h_2$  \cite{Aharony:2002hx}.} $h_1$ and $h_2$,
  \be \label{h12}
 \frac{h_1}{2} \Tr \[ \Phi^1  \Phi^2 \Phi^3 + \Phi^1  \Phi^3 \Phi^2 \]  +\frac{h_2}{2} \Tr \[ (\Phi^1)^2 +  (\Phi^2)^2 +  (\Phi^3)^2  \] \,.
  \ee
 In a neighborhood of free field theory, one confirms  by  inspection of the two-point functions that the only infinite distance direction is associated to the gauge coupling $\tau$, whose metric takes the form (\ref{gaugecuspN=1}). 
Conversely, any point on the $\cN = 4$ SYM fixed line with finite $\tau$ (and $h_1 = h_2 = 0$) is an interior point\footnote{To apply the recipe (\ref{recipe}) when $P$ is a generic point on the $\cN = 4$ fixed line,
one enumerates only the 11 superpotential couplings and takes $G_{\rm global} = SU(3)$, because the $U(1)$ is explicitly broken; this gives again 11-8 = 3 exactly marginal parameters, as it should.} of type (i).
Moving into the directions parametrized by $h_1$ and $h_2$ is a regular perturbation that can be described by conformal perturbation theory. The associated operators are in fact protected, so their two-point functions are independent of $\tau$ to leading order in
$h_1$ and $h_2$.

This example admits a plethora of generalizations. A complete classification of  $\cN =1$ Lagrangian gauge theories with simple gauge group that admit a conformal manifold was recently given in  \cite{Razamat:2020pra}, and many more
cases could be generated with product gauge groups. There are also interesting examples of   theories where the  matter sector that is being conformally gauged contains a strongly-coupled SCFT,
see e.g.~\cite{Razamat:2020gcc} and section 9.2 of \cite{Razamat:2020pra}. All of these gauge theory examples work along similar lines, exhibiting a cusp at $\tau \to i\infty$ with local hyperbolic behavior.

The
 systematic exploration of  $\cN =1$ conformal manifolds  in the search for  dualities 
 has been pioneered in \cite{Razamat:2019vfd, Razamat:2020gcc}. While in its infancy, it has already led to some very curious discoveries.
Let us focus on a  example from  \cite{Razamat:2019vfd} that nicely illustrates the relation between different kinds of symmetry enhancement and distance. The $T_4$ ``trinion" theory is a strongly-coupled $\cN =2$ SCFT with global symmetry $SU(4)^3$ \cite{Gaiotto:2009we}.
While it has no $\cN =2$ exactly marginal deformations (and so it should be viewed an elementary ``matter" building block from the $\cN =2$ viewpoint), it admits an 83-dimensional $\cN =1$ conformal manifold ${\cal M}$. At a generic point of  ${\cal M}$ the
global symmetry is completely broken. A first observation is that   all exactly marginal deformations of $T_4$ are of ``matter" superpotential type. While very special from the viewpoint of global symmetry (and supersymmetry) enhancement,
the $T_4$ point is otherwise a smooth interior point of ${\cal M}$. 
The second much less obvious claim is that ${\cal M}$ admits (at least) another special limiting point, described by  a weakly-coupled fully Lagrangian theory with gauge group $USp(4) \times SU(2)^3$ and 99 chiral multiplets~\cite{Razamat:2019vfd}.
The free gauge theory, with emergent HS symmetry, lies an infinite distance away. 

Our  distance conjecture has some non-trivial implications for the qualitative behavior of the Zamolodchikov distance. We are claiming that either a conformal manifold ${\cal M}$ admits a 
limiting point an infinite distance away with an emergent weakly-coupled (or partially weakly-coupled) dual description, or else it is compact. Take (as just one of many possible examples) the  IR $\cN =1$ SCFT that is obtained from $\cN =4$ SYM as the endpoint of the RG flow triggered by a superpotential mass deformation for one of the three scalar superfields, say $\Phi^3$. 
The IR theory admits a three-dimensional
conformal manifold ${\cal M}_{\rm IR}$, of which a one-dimensional submanifold preserves an $SU(2)$ flavor symmetry (realized in the UV by rotations of $\Phi_1$ and $\Phi_2$). We are predicting that  either ${\cal M}_{\rm IR}$ is compact or there exists  a limiting point with some novel  dual description.\footnote{More elaborate examples of compact $\cN=1$ conformal manifolds have been discussed
in \cite{Buican:2014sfa}.}

\ssec{$d=3$,  $\cN=2$}

Conformal manifolds  of three-dimensional SCFTs  are comparatively much less studied.  The representation theory  for the $d=3$, $\cN =2$ works very similarly to the $d=4$, $\cN=1$ case, and so does
the analysis of \cite{Green:2010da}, leading to the same recipe (\ref{recipe}),  where  the $\chi_i$'s are now interpreted as  chiral primary operators with $\Delta=2$. Accordingly, the status of our conjectures is the same as in $d=4$ SCFT. An important difference is that the Yang-Mills coupling is relevant in three dimensions,
so case (ii) of the previous subsection is never realized.\footnote{As we have already remarked in Section 2, there are many interesting examples of conformal Chern-Simons-matter theories which fall outside the scope of our conjecture because they do not possess a genuine conformal manifold at finite $N$.} In all examples we are aware of, exactly marginal deformations are of the matter superpotential kind. Correspondingly, {\it we expect all $d=3$ conformal manifolds to be compact}. 

A nice example that  has been worked out in detail
 \cite{Baggio:2017mas} is the theory of three chiral superfields with a cubic superpotential.\footnote{See  also \cite{Bachas:2019jaa, Beratto:2020qyk} for recent analyses of $\cN =2$ preserving exactly marginal deformations
of special classes of $\cN =4$ SCFTs.} Its conformal manifold  is an orbifold of ${\rm \bf CP}^1$, endowed with a  metric  that is a certain regular deformation of the Fubini-Study metric
(the deviation from Fubini-Study can be computed perturbatively in $4-\epsilon$ dimensions). While there are special points on ${\cal M}$ with enhanced global symmetry,  none of them has HS symmetry; so the compactness of ${\cal M}$ is 
in concordance with our conjectures. It would be nice to investigate   $d=3$,  $\cN=2$ conformal manifolds more systematically.

\sec{AdS Interpretation}
\label{s3}

Our CFT Distance Conjecture admits a straightforward translation into the language of AdS quantum gravity. We consider an AdS$_{d+1}$ vacuum whose spectrum of excitations includes  massless scalar fields. The bulk moduli space parameterized by scalar vevs is the conformal manifold $\M$ of the dual CFT. Via the holographic dictionary, the scalar two-point functions with conformal boundary conditions define the metric on $\M$. A HS point is holographically dual to an AdS background containing an infinite tower of HS gauge fields. Our conjectures herein relate the spectrum of perturbative HS fields in a fixed AdS background -- that is, with fixed AdS radius in Planck units -- to the metric as follows: approaching a limiting point at infinite distance, a tower of bulk fields of unbounded spin becomes {\it massless}, at a rate exponential in the proper distance. As is perhaps familiar from previous studies of HS holography (e.g. \c{Giombi:2011kc,Aharony:2011jz, Vasiliev:1999ba,Klebanov:2002ja,Gaberdiel:2010pz,Shenker:2011zf, Giombi:2012ms, Didenko:2012tv}) and its embedding in string theory (e.g. \c{Chang:2012kt, Sundborg:2000wp, HaggiMani:2000ru, Sezgin:2002rt, Bianchi:2003wx, Beisert:2003te, Gaberdiel:2015wpo, Gaberdiel:2018rqv, Eberhardt:2018ouy}), the relevant bulk regime is not that of Einstein (super)gravity, whose dual CFTs are far from any cusps on $\M$.\foot{In the present context, we know of no special role played by CFTs with $a\approx c$ and sparse light spectra, which have Einstein gravity duals at strong coupling. It would, however, be interesting to find one.} 

In the context of string theory in AdS$_{D>3}$, our claim is that all infinite distance points in moduli space give rise to massless HS string excitations. For large $N$ CFTs with string duals, this is a statement about the spectrum of classical string theory in an AdS background of large, i.e.~string-scale, curvature; the infinite-dimensional HS gauge symmetry emerges from a tensionless limit, with the spectrum given by that of a strongly-coupled worldsheet sigma model. More generally, the emergent HS symmetry that we are conjecturally associating with infinite distance arguably suggests that the gravity duals have some avatar of stringy structure, in a similar sense as \c{Caron-Huot:2016icg}.  
Also note that our Conjecture I is very natural in the AdS context:  it is a reflection of the implausibility that an Einstein gravity theory can be connected to a Vasiliev-type gravity with infinitely many massless HS fields by a finite field deformation.

In preparation for a more quantitative relation with the DC of the swampland program, let us now write the conjectured exponential behavior \eqr{anomdim} in gravitational variables. Given a spin-$J$ field in AdS with mass $m_J$ approaching zero, we denote by $\a$ the exponential decay rate,
\e{}{m_J := \exp\(-\a\, \widehat{d}(\t,\t')\) \rar 0}
where $\widehat{d}(\t,\t')$ is the (diverging) proper field distance in $(d+1)$-dimensional Planck units. As we will see, this exponent obeys a lower bound. 
 
Near an infinite distance point with metric \eqr{zmetric}, we write the action for the complex scalar modulus $\tau$ as
\e{S0}{S_{\rm bulk}=\ {1\o 2\ell_p^{d-1}}\int d^{d+1}z\sqrt{g}\,\(R - \eta \frac{\p_\mu\t\,\p^\mu\tb}{(\Im \t)^2}+\ldots\)}
$\ell_p$ is the $(d+1)$-dimensional Planck length, $\eta$ is a constant and $z$ are bulk coordinates. We have included the Einstein action to set our conventions for the Planck length. We set the AdS radius to unity. The assertion of the previous section is that as $\text{Im}\,\t \rar \i$, a tower of HS fields becomes massless, $m_J \rar 0$, with scaling \eqr{anomdim}. The AdS/CFT dictionary for spin-$J$ fields implies $m^2_J \propto \g_J$ as $\g_J \rar 0$, so the masses approach zero with parametric scaling $m_J\propto (\Im \t)^{-{1\o 2}}$. Due to the choice of conventions for $\eta$ in \eqr{S0}, we immediately read off the exponent in Planck units as
\e{alphaeq}{\a = \frac{1}{2\sqrt{\eta}}}
We may further relate \eqr{alphaeq} to our field theory parameterization \eqr{zmetric} by using the holographic dictionary to determine the constant of proportionality in $\eta\propto\beta^2$. A metric \eqr{zmetric} defined with respect to a marginal deformation normalized as
\e{}{\delta S_{\rm CFT} = {1\o 4\pi^2}\int d^d x \(\t\, \O_\t(x) + \tb\, \bar\O_\t(x)\)}
yields the following relation obtained by matching two-point functions \c{Freedman:1998tz},
\e{}{{\eta\o 2\ell_p^{d-1}} = {\b^2\o (4\pi^2)^2} { \pi^{d/2}\Gamma(d/2)\o \Gamma(d+1)}\label{eta}~.}
To write $\eta$ in purely CFT terms, we trade $\ell_p^{d-1}$ for $C_T$, the norm of the stress tensor two-point function, using the holographic dictionary (e.g.~\c{Myers:2010tj}), yielding
\e{eta}{{\eta} = {\b^2\o C_T}\, {d+1\o 8\pi^4(d-1)}~.}
By writing the distance in Planck units, we have computed $\a$ in units of the central charge $C_T$. 

Let us make the following brief remark. One may find it natural to use CFT conventions that rescale the distance by a factor of $\sqrt{C_T}$. In that case, $\b^2/C_T$ instead of $\b^2$ parameterizes the asymptotic behavior \eqr{zmetric}. Though the strongest motivation for this choice comes from natural (Planck) units in holography, the quantity $C_T$ is simply the canonical stress tensor normalization and is not inherently holographic. This suggests that the normalization of the Zamolodchikov metric by the stress tensor two-point function is an interesting quantity even at finite $C_T$.

We pause to note that AdS vacua in quantum gravity often appear together with a compact space, AdS$_{d+1}\times X_{D-d}$ (which is also a consequence of a generalized DC of the Swampland in the AdS context \cite{Lust:2019zwm}). One may wonder how sensitive $\a$ is to whether we measure the field distance in $(D+1)$- or $(d+1)$-dimensional Planck units. They are related as $\ell_{p,D+1}^{D-1}=\ell_{p}^{d-1} \text{Vol}(X)$. Therefore, if there is no scale separation between the AdS and $X$ radii, then $\a$ changes only by order one numerical factors inherited from the volume $\text{Vol}(X)$. The same applies if $X$ is a direct product space with only a subset of its dimensions admitting scale separation.

\ssec{A lower bound for the exponential rate}\label{s31}
The result \eqr{alphaeq} shows that $\a$ admits a lower bound: the scalar kinetic term in \eqr{S0} must have finite normalization $\eta$ relative to the gravitational term, since both are normalized with respect to the Planck length which defines the shortest length scale in quantum gravity. A lower bound is interesting because it implies in turn an upper bound on the scalar field range that can be travelled in the moduli space before the quantum gravity breakdown induced by HS fields. However, there is no upper bound, as the scalar kinetic term may be parametrically suppressed (as we will see later). 

When $\t$ is a complexified gauge coupling in $d=4$, we earlier derived $\b^2=24\,\text{dim}\,G$. Trading $C_T = {40\o \pi^4} c$ for convenience, where $c$ is the $c$-type central charge defined in the conventions of \c{Myers:2010tj}, plugging into \eqref{eta} yields 
\e{alphaeq2}{\a = \sqrt{2c\o \text{dim}\,G}  \qquad (d=4)}
Thus, $\a$ is minimized by the CFT with the smallest value of $c$ relative to ${\rm dim}\,G$. For application to $\cN=2$ SCFTs, recalling that the $c$-type central charge of a single $\cN = 2$ vector multiplet is $c_{\rm v} =1/6$, one may write this as $\a = \sqrt{c/3 c_{\rm v}}$. Analogously, for $\cN=1$ SCFTs, we get  $\a = \sqrt{c/4c_{\rm v}}$ where $c_{\rm v} =1/8$ for a single $\cN = 1$ vector multiplet. There seems to be no upper bound for $\alpha$, as we can in principle select a subsector $c_{\rm v}\ll c$. On the other hand, for the case of vanishing gauge couplings, a lower bound on $\a$ follows from $c> c_{\rm v}$. Noting that a gauge theory with $c=c_{\rm v}$ is not conformal, we conclude that $\alpha$ satisfies the following strict inequalities,
\beq
\label{bounds}
\alpha>\frac{1}{\sqrt{3}} \qquad (d=4,\ \cN=2)\quad ; \quad \alpha>\frac{1}{\sqrt{4}} \qquad (d=4,\ \cN=1)
\eeq
This begs the question of how much these bounds can be improved. We now prove that in several classes of large $c$ SCFTs with $\cN=1$ and $\cN=2$ supersymmetry listed below, $\a$ obeys the stronger bound
\e{alphabound}{\a \geq {1\o \sqrt{2}} \quad (d=4)}
Moreover, this bound is saturated by $\cN=4$ SYM, for which $c={{\rm dim} \, G\o 4}$.  

\begin{itemize}

\item {\it $\cN=2$ SCFTs with simple gauge group:} A complete classification of $d=4, \cN=2$ Lagrangian gauge theories was presented in \cite{Bhardwaj:2013qia}. Extracting those which admit a large $N$ limit, we compute the exponent $\alpha$ using \eqref{alphaeq2} and present the results in Table \ref{table}. All of them satisfy \eqref{alphabound}.\foot{We also computed \eqr{alphaeq2} for all of the $\cN=2$ theories in \cite{Bhardwaj:2013qia} with finite $N$. This calculation is motivated by the discussion below \eqr{eta}. The result is that all such cases satisfy the bound \eqref{alphabound} with the exception of $USp(4)$ with a half hypermultiplet in the 16 representation. Amusingly, this theory has the property, unusual among Lagrangian SCFTs, that $a>c$. We conclude that the stricter bound \eqr{alphabound} does not hold away from large $N$. We leave open the interesting question of the optimal bound at finite $N$.}

\begin{table}[t]
\centering
\renewcommand{\arraystretch}{1.4}
\begin{tabular}{!{\color{black}\vrule}c!{\color{black}\vrule}l!{\color{black}\vrule}c!{\color{black}\vrule}c!{\color{black}\vrule}} 
\hline
\rowcolor[rgb]{0.753,0.753,0.753} $G$        & Hypermultiplets & $c$      & $\alpha$                                 \\ 
\hline
{\cellcolor[rgb]{0.753,0.753,0.753}}$SU(N)$  & $2N$ \fund           & $\frac16 (2N^2-1)$ & $\sqrt{\frac{2}{3}}$  \\ 
\hline
{\cellcolor[rgb]{0.753,0.753,0.753}}$SU(N)$  & $1$ \asym, $N+2$ \fund           & $\frac1{24}(7N^2+3N-4)$ & $\sqrt{\frac{7}{12}}$  \\ 
\hline
{\cellcolor[rgb]{0.753,0.753,0.753}}$SU(N)$  & $2$ \asym, $4$ \fund           & $\frac1{12}(3N^2+3N-2)$ & $\frac{1}{\sqrt{2}}$  \\ 
\hline
{\cellcolor[rgb]{0.753,0.753,0.753}}$SU(N)$  & $1$ \asym, $N-2$ \fund           & $\frac1{24}(7N^2-3N-4)$ & $\sqrt{\frac{7}{12}}$  \\ 
\hline
{\cellcolor[rgb]{0.753,0.753,0.753}}$SU(N)$  & $1$ \sym, $1$ \asym           & $\frac{1}{12}(3N^2-2)$ & $\frac{1}{\sqrt{2}}$  \\ 
\hline
{\cellcolor[rgb]{0.753,0.753,0.753}}$USp(2N)$  &  $4N+4$ $\frac12$ \fund               &    $ \frac16 N(4N+3)$       &        $\sqrt{2\o3}$                                                   \\ 
\hline
{\cellcolor[rgb]{0.753,0.753,0.753}}$USp(2N)$ &      $1$ \asym, $4$ \fund           &                  $\frac1{12}(6N^2+9N-1)$      &   $\frac{1}{\sqrt{2}}$                                           \\ 
\hline
{\cellcolor[rgb]{0.753,0.753,0.753}}$SO(N)$  & $N-2$ {\bf vect}           & $\frac1{12}N(2N-3)$ &$ \sqrt{\frac{2}{3}}$  \\ 
\hline
\end{tabular}
\caption{Computation of $\alpha$ for $d=4, \cN=2$ SCFTs with simple gauge group, to leading order in large $N$. We list all entries of the classification of \cite{Bhardwaj:2013qia} that admit large $N$ limits, in the notation of \cite{Bhardwaj:2013qia}.}
\label{table}
\end{table}

\item {\it $\cN=1$ SCFTs with simple gauge group:} A complete classification of $d=4, \cN=1$ Lagrangian conformal manifolds with simple gauge groups was presented in \cite{Razamat:2020pra}. Extracting those which admit a large $N$ limit, we compute the exponent $\alpha$ using \eqref{alphaeq2} and present the results in Table \ref{tableN1}. All of them satisfy \eqref{alphabound}.

\begin{table}[t]
\renewcommand{\arraystretch}{1.4}
\centering
\begin{tabular}{!{\color{black}\vrule}c!{\color{black}\vrule}l!{\color{black}\vrule}c!{\color{black}\vrule}c!{\color{black}\vrule}} 
\hline
\rowcolor[rgb]{0.753,0.753,0.753} $G$        & Theory & $c$      & $\alpha$                                 \\ 
\hline
{\cellcolor[rgb]{0.753,0.753,0.753}}$SU(N)$  & Table 2, \#1           & $\frac{1}{24}(7N^2-5)$ & $\sqrt{7\o 12}$  \\ 
\hline
{\cellcolor[rgb]{0.753,0.753,0.753}}$SU(N)$  & Table 2, \#5        & $\frac{1}{24}(6N^2+3N-5)$ & ${1\o\sqrt{2}}$  \\ 
\hline
{\cellcolor[rgb]{0.753,0.753,0.753}}$SU(N)$  & Table 3, \#4        & $\frac{1}{24}(7N^2-4)$ & $\sqrt{7\o 12}$  \\ 
\hline
{\cellcolor[rgb]{0.753,0.753,0.753}}$SU(N)$  & Table 5, \#4        & $\frac{1}{24}(8N^2-3)$ & $\sqrt{2\o 3}$  \\ 
\hline
{\cellcolor[rgb]{0.753,0.753,0.753}}$USp(2N)$  & Table 12, \#1        & $\frac{1}{24}(14N^2+15N-1)$ & $\sqrt{7\o 12}$  \\ 
\hline
{\cellcolor[rgb]{0.753,0.753,0.753}}$USp(2N)$  & Table 13, \#9        & $\frac{1}{8}(4N^2+8N-1)$ & $\frac{1}{\sqrt{ 2}}$  \\ 
\hline
{\cellcolor[rgb]{0.753,0.753,0.753}}$USp(2N)$  & Table 13, \#10        & $\frac{1}{24}(14N^2+21N-2)$ & $\sqrt{7\o 12}$  \\ 
\hline
{\cellcolor[rgb]{0.753,0.753,0.753}}$SO(N)$  & Table 18, \#1        & $\frac{1}{48}(7N^2-21N-4)$ & $\sqrt{7\o 12}$  \\ 
\hline
{\cellcolor[rgb]{0.753,0.753,0.753}}$SO(N)$  & Table 18, \#2        & $\frac{1}{48}(7N^2-15N-2)$ & $\sqrt{7\o 12}$  \\ 
\hline
{\cellcolor[rgb]{0.753,0.753,0.753}}$SO(N)$  & Table 18, \#3        & $\frac{1}{24}(4N^2-9N-1)$ & $\sqrt{2\o 3}$  \\ 
\hline
\end{tabular}
\caption{Computation of $\alpha$ for $d=4, \cN=1$ SCFTs with simple gauge group, to leading order in large $N$. We list all entries of the classification of \cite{Razamat:2020pra} that admit large $N$ limits, referring to the table entry in \cite{Razamat:2020pra}.}
\label{tableN1}
\end{table}

\item {\it $\cN=1$ SCFTs with AdS$_5\times SE_5$ duals:} Consider $\cN=1$ SCFTs with conformal manifolds and $a\approx c$ at large $N$. Some of these  SCFTs admit a supergravity description at strong coupling. A well-studied class of supergravity backgrounds dual to $\cN=1$ SCFTs are AdS$_5\times SE_5$ solutions of type IIB, where $SE_5$ is a Sasaki-Einstein 5-manifold. Consider a theory in this class, call it $\mathcal{T}_*$, with a weakly-coupled fixed point with $G$-valued gauge fields. Compared to $\cN=4$ SYM with the same gauge group $G$, $\mathcal{T}_*$  will have $\a_* > \a_{\cN=4}$ if and only if $c_* > c_{\cN=4}$. Due to supersymmetry, $c$ is not renormalized and can hence be computed in supergravity. The holographic dictionary \c{Gubser:1998vd} fixes the central charge $c$ to be inversely proportional to the volume of the Sasaki-Einstein $SE_5$ manifold,
\e{cvol}{c\propto {1\o {\rm Vol}(SE_5)}}
It is an old result in the theory of Einstein manifolds (not necessarily Sasakian), due to Bishop \c{bishop1964geometry}, that the volume of a closed Einstein $n$-manifold is bounded above by that of the unit round $n$-sphere, with saturation only for the sphere. Applied to our case,
\e{}{{{\rm Vol}(SE_5)\o {\rm Vol}(S^5)}< 1\,,\quad SE_5 \neq S^5\,,}
a statement which is readily checked for examples with known metrics, e.g.~$L^{pqr}$ spaces \c{Cvetic2005}. Therefore, \eqr{cvol} implies $\a_* > \a_{\cN=4} = {1\o \sqrt{2}}$. 

\end{itemize}

In contrast to these lower bounds, a simple example in which $\a$ can become arbitrarily {\it large} is the case of $d=4, \cN=2$ class $\mathcal{S}$ theories, reviewed in e.g.~\c{Beem:2014rza}. Associated to a class $\mathcal{S}$ theory is, among other data, a simply-laced Lie algebra and a (possibly punctured) genus $g$ Riemann surface. Focusing on the $A_N$ series for concreteness, the central charge is $c\approx {N^3\o 3}(g-1)$ to leading order in large $N$. Since ${\rm dim}\,G \sim N^2$, \eqr{alphaeq2} will scale with $N$. Explicitly,
\beq
\alpha\approx\sqrt{\frac{2N(g-1)}{3k}\,,}
\eeq
where $k$ is again the number of complexified gauge couplings sent to zero simultaneously, bounded by the genus as $k\leq g$. The $\a\sim \sqrt{N}$ scaling for fixed $g$ and large $N$, a somewhat surprising result, highlights a difference between the CFT Distance Conjecture and expectations from the DC in flat space, in which $\a$ is $O(1)$.\foot{In \c{Tachikawa:2017aux}, the K\"ahler class of the Zamolodchikov metric of class $\mathcal{S}$ theories was computed from an anomaly perspective. However, in order to extract the metric from those computations, one must avoid degeneration limits in which the K\"ahler class is ill-defined. Our computation of $\a$ is relevant precisely in the degeneration limit, so there is no comparison to be made. The authors of \c{Tachikawa:2017aux} also compute the Zamolodchikov metric from supergravity, but that is likewise not at infinite distance. We thank Yuji Tachikawa for helpful correspondence.}

\ssec{Comparison with the DC in the swampland program}\label{s32}

As explained in the introduction, the AdS formulation of the CFT Distance Conjecture makes contact with the Distance Conjecture (DC) in the swampland program. In this section, we will delve into this comparison, although the reader should keep in mind that the CFT Distance Conjecture applies regardless of whether the CFT has a semiclassical dual AdS description. Even though it pertains to families of CFTs with fixed central charge, which is incompatible with taking the flat space limit of AdS \c{Polchinski:1999ry}, it is natural to expect that the DC applies also in the AdS background.  Higher-spin symmetry plays an essential role in our conjectures, which assume $d>2$. However, similar conjectures \c{Kontsevich:2000yf, KS,Acharya:2006zw} in $d=2$ involve instead the scalar gap. Hence,  as noted in Section \ref{s2}, it is possible to unify both CFT conjectures into a weaker statement valid for $d\geq 2$ involving infinite towers of operators at infinite distance which are dual to massless fields in AdS, without referring to the spin. This more universal but weaker statement maps directly to the DC in the swampland program, while the additional constraints on the spin can be understood as stronger refinements of the latter for the specific case of AdS. Indeed, from the perspective of the DC in the Swampland program, for large $N$ CFTs with an Einstein gravity dual the HS aspect of our conjectures is natural: the emergence of a tower of light states in the DC for AdS$_{d+1}$ with $d>2$ would imply a tower of vanishing anomalous dimensions in the dual CFT for HS states, since an accumulation of spin $J\leq 2$ operators in $d>2$ is forbidden by the expected finiteness of the CFT partition function even in the infinite distance limit.

In the context of the DC in the swampland program, one of the most important open questions is to understand the properties, and range of possible values, of the exponential decay rate of the tower. Progress in this direction has been performed in the context of Calabi-Yau (CY) flat space compactifications \cite{Grimm:2018ohb,Grimm_2019,Corvilain:2018lgw,lee2019emergent,Gendler:2020dfp}, where the exponential decay rate is parametrized by the monodromy properties of the large field limit and roughly indicates  how much of the space is getting decompactified. There is some similarity with the result for the exponent $\a$ for the gauge case in \eqr{alphaeq2}, which parametrizes what portion of the theory decouples at infinite distance. In the CY case, a lower bound on the exponent corresponds to decompactifying all dimensions; in AdS, the lower bound corresponds to the full CFT becoming free. Furthermore, in the CY case, the exponent also has a maximum, corresponding to equi-dimensional string perturbative limits. Thus, the upper bound is possible only because the maximum number of total dimensions is bounded for effective field theories arising from (sub)critical string theory and M-theory compactifications. On the contrary, $\a$ in AdS has no upper bound, so the tower can get light parametrically fast. Of course, in both AdS and flat space, these constraints disappear when decoupling gravity; this is as expected from experience in the swampland.

Let us analyse in more detail the specific value of the lower bound for $\alpha$. In the context of the DC in quantum gravity, determining precisely this lower bound is one of the most important aspects of the conjecture. As mentioned in the introduction, the TCC motivates a lower bound for the DC exponent given either by $\alpha_{TCC}\geq \frac{2}{D\sqrt{(D-1)(D-2)}}$ \cite{Bedroya:2019snp} or $\alpha_{TCC}\geq \frac{1}{\sqrt{(D-1)(D-2)}}$ \cite{Andriot:2020lea,Bedroya:2020rmd} depending on the specific relation between the mass of the tower and the scalar potential, where $D$ is the spacetime dimension. This yields  $\alpha_{TCC}\geq \frac1{5\sqrt{3}}$ or $\alpha_{TCC}\geq \frac1{\sqrt{12}}$ for $D=5$, respectively. In the previous subsection we collected evidence for a universal lower bound of $\alpha \geq {1\o \sqrt{2}}$ for the case of vanishing gauge couplings, which we proved to hold for some prominent classes of SCFTs. This value is larger than the TCC prediction. It is also larger than the bound obtained for Calabi-Yau $CY_3$ compactifications, namely $\alpha_{CY_3}\geq \frac1{\sqrt{6}}$ \cite{Grimm:2018ohb,Gendler:2020dfp}.\footnote{For theories with eight supercharges arising from $CY_n$ compactifications, it was found in \cite{Grimm:2018ohb,Gendler:2020dfp} a universal bound $\alpha_{CY}\geq \frac1{\sqrt{2n}}$ with $n= $ dim$_\mathbb{C}(CY)$ for towers of BPS states.} Amusingly, for the case of simple gauge groups, we found in the previous section that $\alpha$ can only take three values: $\alpha= \sqrt{6/12}, \sqrt{7/12}, \sqrt{8/12}$, whose squares are all multiples of $1/12$ which is precisely the above minimum value predicted by the TCC.

How does the swampland DC inform the CFT Distance Conjecture? Conjecture III and the exponential decay of HS anomalous dimensions \eqr{anomdim} were shown here to hold in the gauge case, but are conjectural more generally. On the other hand, exponential decay is the central feature of the swampland DC: not only has it been tested in a wider variety of top-down models, but its origins (in certain supersymmetric cases) have been more rigorously tied to certain theorems of limiting mixed Hodge structures \cite{Grimm:2018ohb,Grimm_2019,Cecotti:2020rjq,Grimm:2020cda}. We take this as an independent suggestion that the Zamolodchikov metric asymptoting to infinite distance is indeed locally hyperbolic. In fact, it would be interesting to check whether the same mathematical techniques based on Hodge theory used in the context of CY spaces might also be used to prove our Conjectures I and III for conformal manifolds in Section \ref{s2}. We leave this for future work.

Conversely, the CFT Distance Conjecture lends support for stronger versions of the DC that specify the nature of the tower of states. In particular, the CFT Distance Conjecture requires the universal presence of HS fields furnishing an emergent infinite-dimensional gauge symmetry at infinite distance in $d>2$. This suggests similar properties at infinite distance in flat space quantum gravity. It also resonates with the proposal in \cite{Lanza:2020qmt}, for which every infinite field distance limit can be identified with the RG flow endpoint of codimension-two objects universally present at these limits (i.e. particles in $D=d+1=3$ and extended objects containing HS fields in $D>3$), and with the proposal in \cite{Gendler:2020dfp} that any infinite distance limit corresponds to the weak coupling limit of some $p$-form gauge field. The CFT Distance Conjecture lends a different conceptual motivation for these. In \cite{lee2019emergent} the Emergent String Conjecture was proposed, which claims that any infinite distance limit should either correspond to a decompactification limit or a limit in which a critical weakly-coupled string becomes tensionless with respect to the Planck mass.  Our conjecture in $d>2$ is consistent with the latter case, as the string spectra contain towers of HS fields, but we are making the prediction that at fixed AdS curvature, there can be no (even partial) decompactification of $X$ in AdS$_{D>3} \times X$: in particular, assuming that decompactification would lead to an infinite tower of massless scalars, our conjecture asserts that the massless modes in AdS must instead be HS fields. Note that our HS conjecture in the context of holographic dual AdS$_{D>3}\times X$ implies that the scale of no part of $X$ can be decoupled from that of AdS and thus  leads to a refined version of the absence of AdS scale separation \cite{Lust:2019zwm}, which in principle allows for part of $X$ to decompactify (as is the case in AdS$_3$ examples).\footnote{Another possibility, which we think is unlikely to be the case, is for our HS conjecture to be incorrect without invalidating the DC in the Swampland context by getting a partial decompactification of $X$ leading to a light tower of KK modes comprised of massless $J=0$ modes in this limit.}

\ssec{Application to some AdS/CFT dual pairs}\label{s33}
We close this section with some phenomenology for $d=4$ SCFTs at large central charge. We compute $\a$ from CFT and make some comments about the AdS dual in the relevant regime. To be clear, AdS/CFT guarantees that the bulk and boundary HS spectra will agree everywhere on $\M$. In general, one cannot use a supergravity approximation (if it exists), which does not lie at infinite distance, to compute the spectrum and the metric near a HS point. For the special case of $\cN=4$ SYM at large $N$, we explain the sense in which one can actually match to a supergravity computation.

\subsubsection{$\cN=4$ $SU(N)$ super-Yang-Mills}
As derived earlier,  
\beq
\alpha=\frac{1}{\sqrt{2}}
\eeq
There is substantial literature on the local operator spectrum of free $\cN=4$ SYM \c{Sundborg:2000wp, HaggiMani:2000ru, Sezgin:2002rt, Beisert:2003te, Sundborg:1999ue, Aharony:2003sx}, including the algebra of HS conserved currents \c{Beisert:2004di}. The bulk dual is type IIB string theory on AdS$_5\times S^5$. In the regime of large (but finite) $N$, small coupling $g_s \sim g_{\rm YM}^2 \ll 1$ and small 't Hooft coupling $\lambda = g_{\rm YM}^2 N \ll1$, a good approximation to the bulk theory is given by classical
string theory in a highly curved (string size) geometry. It is believed that in this regime, the holographic dictionary is modified such that $L\sim \ell_s$ corresponds to $\lambda = 0$. (For a compelling case, see \c{Sezgin:2002rt}; analogous claims in AdS$_3\times S^3 \times T^4$ were firmly established in \c{Eberhardt:2018ouy}.) Nevertheless, due to the maximal supersymmetry, the exponent $\a$ may be computed in the {\it supergravity} regime. Maximal supersymmetry implies that the metric is in fact given by the Poincar\'e metric over {\it all} of moduli space, and one needs only determine the overall constant $\b^2$. We are free to compute $\b$ in type IIB supergravity. From e.g.~\c{DHoker:2002nbb}, the graviton-axio-dilaton part of the type IIB supergravity action in Einstein frame takes the form \eqr{S0} with $\eta=1/2$, in agreement with the CFT result above. 

One may also perform $\mathbb{Z}_n$ orbifolds of $\cN=4$ SYM that preserve $\cN=2$ supersymmetry, dual to type IIB string theory on AdS$_5\times S^5/\mathbb{Z}_n$, with central charge $c = n\,\text{dim}\, G /4$. The gauge group is $SU(N)^n$, so the moduli space of this conformal manifold is parameterized by one complexified gauge coupling for each gauge factor. If we take the weak coupling limit of $k$ gauge couplings simultaneously, where $k=1,\dots, n$, we get 
\e{}{\alpha=\sqrt{\frac{n}{2k}}}
There is no upper bound on the rank $n$ in the CFT, though the classical gravity approximation breaks down when Vol$(S^5/\mathbb{Z}_n) = {\rm Vol}(S^5)/n$ is of order the Planck scale.

\subsubsection{$SU(N)$ SQCD with $N_f=2N$}

We next consider $SU(N)$ SQCD with $N_f=2N$ at large $N$, which preserves $\cN=2$ supersymmetry. From the first entry in Table \ref{table},
\e{sqcdalpha}{\alpha\approx \sqrt{\frac23}}
The bulk dual of this SCFT was explored in \c{Gadde:2009dj}. $\cN=2$ supersymmetry is not powerful enough to constrain the metric to be globally hyperbolic, unlike $\cN=4$ SYM. In order to faithfully match \eqr{sqcdalpha} to a bulk computation, one must compute the HS spectrum in the highly stringy regime of large curvature. (See \c{Gadde:2009dj} for some comments and speculations about the string-scale geometry.) We note that the bulk dual of SQCD has no supergravity regime anywhere on $\M$, as $a\neq c$ at large $N$. 

\sssec{$d=4, \cN=1$ $\b$-deformed CFTs}

The result \eqr{gaugecuspN=1} gives the asymptotic behavior of the metric in a $d=4, \cN=1$ SCFT as a gauge coupling vanishes. One class of $\cN=1$ SCFTs with well-studied holographic duals are $\b$-deformations, which preserve $\cN=1$ supersymmetry and a $U(1)\times U(1)$ flavor symmetry. The result \eqr{gaugecuspN=1} is invariant under $\b$-deformation, which preserves the gauge group $G$. AdS vacua for $\b$-deformed SCFTs were constructed in supergravity by performing an $SL(2,\mathbb{R})$ transformation of the original solutions \cite{Lunin:2005jy}. By inspection of the actions given  in \c{Lunin:2005jy}, one can notice that the directions associated to superpotential couplings are compact while the one associated to the gauge coupling is non-compact, in harmony with the discussion in Section \ref{N=1}. Focusing on the $\b$-deformation of $\cN=4$ SYM for concreteness, one observes that the purely axio-dilaton action given in \c{Lunin:2005jy} is identical to the one in AdS$_5\times S^5$, before the $SL(2,\mathbb{R})$ transformation. This would suggest the same value of $\alpha$ as before the $\beta$-deformation if the leading term of the bulk field metric could be trusted at weak coupling; however, {\it a priori} this is not the case. It would be interesting to determine the fate of the backgrounds of \c{Lunin:2005jy} and their associated string spectra as one interpolates between supergravity and string theory, and whether non-renormalization theorems exist. 

\sec*{Acknowledgments}
We wish to thank Chris Beem, Jan de Boer, Sergio Cecotti, Tristan Collins, Daniel Jafferis, Zohar Komargodski, Jacob McNamara, Carlo Meneghelli, Miguel Montero, Kyriakos Papadodimas, Shlomo Razamat, Yuji Tachikawa, Arnav Tripathy and Shing-Tung Yau for helpful discussions.

 E.P. is supported in part by the World Premier International Research Center Initiative, MEXT, Japan, and by the U.S. Department of Energy, Office of Science, Office of High Energy Physics, under Award Number DE-SC0011632. The research of L.R. is supported in part by NSF grant No. PHY-1915093.  The research of C.V. and I.V. was supported in part by a grant from the Simons Foundation (602883, CV). The research of  C.V. was also supported by the National Science Foundation under Grant No. NSF PHY-2013858. We thank the KITP for hospitality during the course of this work, which was supported in part by the National Science Foundation under Grant No. NSF PHY-1748958.

\bibliographystyle{utphys}
\bibliography{SDCbiblio}

\providecommand{\href}[2]{#2}\begingroup\raggedright\begin{thebibliography}{100}

\bibitem{Simmons-Duffin:2016gjk}
D.~Simmons-Duffin, \href{http://dx.doi.org/10.1142/9789813149441_0001}{``{The
  Conformal Bootstrap},''} in {\em {Proceedings, Theoretical Advanced Study
  Institute in Elementary Particle Physics: New Frontiers in Fields and Strings
  (TASI 2015): Boulder, CO, USA, June 1-26, 2015}}, pp.~1--74.
\newblock 2017.
\newblock
\href{http://arxiv.org/abs/1602.07982}{{\ttfamily arXiv:1602.07982 [hep-th]}}.
\newblock

\bibitem{Poland:2018epd}
D.~Poland, S.~Rychkov, and A.~Vichi, ``{The Conformal Bootstrap: Theory,
  Numerical Techniques, and Applications},''
\href{http://arxiv.org/abs/1805.04405}{{\ttfamily arXiv:1805.04405 [hep-th]}}.

\bibitem{Brennan:2017rbf}
T.~D. Brennan, F.~Carta, and C.~Vafa, ``{The String Landscape, the Swampland,
  and the Missing Corner},'' \href{http://dx.doi.org/10.22323/1.305.0015}{{\em
  PoS} {\bfseries TASI2017} (2017) 015},
  \href{http://arxiv.org/abs/1711.00864}{{\ttfamily arXiv:1711.00864
  [hep-th]}}.

\bibitem{Palti:2019pca}
E.~Palti, ``{The Swampland: Introduction and Review},''
  \href{http://dx.doi.org/10.1002/prop.201900037}{{\em Fortsch. Phys.}
  {\bfseries 67} no.~6, (2019) 1900037},
  \href{http://arxiv.org/abs/1903.06239}{{\ttfamily arXiv:1903.06239
  [hep-th]}}.

\bibitem{Nakayama:2015hga}
Y.~Nakayama and Y.~Nomura, ``{Weak gravity conjecture in the AdS/CFT
  correspondence},'' \href{http://dx.doi.org/10.1103/PhysRevD.92.126006}{{\em
  Phys. Rev. D} {\bfseries 92} no.~12, (2015) 126006},
  \href{http://arxiv.org/abs/1509.01647}{{\ttfamily arXiv:1509.01647
  [hep-th]}}.

\bibitem{Harlow:2015lma}
D.~Harlow, ``{Wormholes, Emergent Gauge Fields, and the Weak Gravity
  Conjecture},'' \href{http://dx.doi.org/10.1007/JHEP01(2016)122}{{\em JHEP}
  {\bfseries 01} (2016) 122}, \href{http://arxiv.org/abs/1510.07911}{{\ttfamily
  arXiv:1510.07911 [hep-th]}}.

\bibitem{Benjamin:2016fhe}
N.~Benjamin, E.~Dyer, A.~L. Fitzpatrick, and S.~Kachru, ``{Universal Bounds on
  Charged States in 2d CFT and 3d Gravity},''
  \href{http://dx.doi.org/10.1007/JHEP08(2016)041}{{\em JHEP} {\bfseries 08}
  (2016) 041}, \href{http://arxiv.org/abs/1603.09745}{{\ttfamily
  arXiv:1603.09745 [hep-th]}}.

\bibitem{Montero:2016tif}
M.~Montero, G.~Shiu, and P.~Soler, ``{The Weak Gravity Conjecture in three
  dimensions},'' \href{http://dx.doi.org/10.1007/JHEP10(2016)159}{{\em JHEP}
  {\bfseries 10} (2016) 159}, \href{http://arxiv.org/abs/1606.08438}{{\ttfamily
  arXiv:1606.08438 [hep-th]}}.

\bibitem{Heidenreich:2016aqi}
B.~Heidenreich, M.~Reece, and T.~Rudelius, ``{Evidence for a sublattice weak
  gravity conjecture},'' \href{http://dx.doi.org/10.1007/JHEP08(2017)025}{{\em
  JHEP} {\bfseries 08} (2017) 025},
  \href{http://arxiv.org/abs/1606.08437}{{\ttfamily arXiv:1606.08437
  [hep-th]}}.

\bibitem{Montero:2017mdq}
M.~Montero, ``{Are tiny gauge couplings out of the Swampland?},''
  \href{http://dx.doi.org/10.1007/JHEP10(2017)208}{{\em JHEP} {\bfseries 10}
  (2017) 208}, \href{http://arxiv.org/abs/1708.02249}{{\ttfamily
  arXiv:1708.02249 [hep-th]}}.

\bibitem{Harlow:2018tng}
D.~Harlow and H.~Ooguri, ``{Symmetries in quantum field theory and quantum
  gravity},'' \href{http://arxiv.org/abs/1810.05338}{{\ttfamily
  arXiv:1810.05338 [hep-th]}}.

\bibitem{Bae:2018qym}
J.-B. Bae, S.~Lee, and J.~Song, ``{Modular Constraints on Superconformal Field
  Theories},'' \href{http://dx.doi.org/10.1007/JHEP01(2019)209}{{\em JHEP}
  {\bfseries 01} (2019) 209}, \href{http://arxiv.org/abs/1811.00976}{{\ttfamily
  arXiv:1811.00976 [hep-th]}}.

\bibitem{Harlow:2018jwu}
D.~Harlow and H.~Ooguri, ``{Constraints on Symmetries from Holography},''
  \href{http://dx.doi.org/10.1103/PhysRevLett.122.191601}{{\em Phys. Rev.
  Lett.} {\bfseries 122} no.~19, (2019) 191601},
  \href{http://arxiv.org/abs/1810.05337}{{\ttfamily arXiv:1810.05337
  [hep-th]}}.

\bibitem{Lin:2019kpn}
Y.-H. Lin and S.-H. Shao, ``{Anomalies and Bounds on Charged Operators},''
  \href{http://dx.doi.org/10.1103/PhysRevD.100.025013}{{\em Phys. Rev. D}
  {\bfseries 100} no.~2, (2019) 025013},
  \href{http://arxiv.org/abs/1904.04833}{{\ttfamily arXiv:1904.04833
  [hep-th]}}.

\bibitem{Montero:2018fns}
M.~Montero, ``{A Holographic Derivation of the Weak Gravity Conjecture},''
  \href{http://dx.doi.org/10.1007/JHEP03(2019)157}{{\em JHEP} {\bfseries 03}
  (2019) 157}, \href{http://arxiv.org/abs/1812.03978}{{\ttfamily
  arXiv:1812.03978 [hep-th]}}.

\bibitem{Conlon:2018vov}
J.~P. Conlon and F.~Quevedo, ``{Putting the Boot into the Swampland},''
  \href{http://dx.doi.org/10.1007/JHEP03(2019)005}{{\em JHEP} {\bfseries 03}
  (2019) 005}, \href{http://arxiv.org/abs/1811.06276}{{\ttfamily
  arXiv:1811.06276 [hep-th]}}.

\bibitem{Conlon:2020wmc}
J.~P. Conlon and F.~Revello, ``{Moduli Stabilisation and the Holographic
  Swampland},'' \href{http://arxiv.org/abs/2006.01021}{{\ttfamily
  arXiv:2006.01021 [hep-th]}}.

\bibitem{Ooguri:2020sua}
H.~Ooguri and T.~Takayanagi, ``{Cobordism Conjecture in AdS},''
  \href{http://arxiv.org/abs/2006.13953}{{\ttfamily arXiv:2006.13953
  [hep-th]}}.

\bibitem{Agarwal:2020pol}
P.~Agarwal, K.-H. Lee, and J.~Song, ``{Classification of large N superconformal
  gauge theories with a dense spectrum},''
  \href{http://arxiv.org/abs/2007.16165}{{\ttfamily arXiv:2007.16165
  [hep-th]}}.

\bibitem{Ooguri:2006in}
H.~Ooguri and C.~Vafa, ``{On the Geometry of the String Landscape and the
  Swampland},'' \href{http://dx.doi.org/10.1016/j.nuclphysb.2006.10.033}{{\em
  Nucl. Phys.} {\bfseries B766} (2007) 21--33},
\href{http://arxiv.org/abs/hep-th/0605264}{{\ttfamily arXiv:hep-th/0605264
  [hep-th]}}.

\bibitem{Klaewer_2017}
D.~Klaewer and E.~Palti, ``{Super-Planckian Spatial Field Variations and
  Quantum Gravity},'' \href{http://dx.doi.org/10.1007/JHEP01(2017)088}{{\em
  JHEP} {\bfseries 01} (2017) 088},
\href{http://arxiv.org/abs/1610.00010}{{\ttfamily arXiv:1610.00010 [hep-th]}}.

\bibitem{Grimm:2018ohb}
T.~W. Grimm, E.~Palti, and I.~Valenzuela, ``{Infinite Distances in Field Space
  and Massless Towers of States},''
  \href{http://dx.doi.org/10.1007/JHEP08(2018)143}{{\em JHEP} {\bfseries 08}
  (2018) 143},
\href{http://arxiv.org/abs/1802.08264}{{\ttfamily arXiv:1802.08264 [hep-th]}}.

\bibitem{Lee:2018urn}
S.-J. Lee, W.~Lerche, and T.~Weigand, ``{Tensionless Strings and the Weak
  Gravity Conjecture},'' \href{http://dx.doi.org/10.1007/JHEP10(2018)164}{{\em
  JHEP} {\bfseries 10} (2018) 164},
\href{http://arxiv.org/abs/1808.05958}{{\ttfamily arXiv:1808.05958 [hep-th]}}.

\bibitem{Lee_2019}
S.-J. Lee, W.~Lerche, and T.~Weigand, ``{A Stringy Test of the Scalar Weak
  Gravity Conjecture},''
  \href{http://dx.doi.org/10.1016/j.nuclphysb.2018.11.001}{{\em Nucl. Phys.}
  {\bfseries B938} (2019) 321--350},
\href{http://arxiv.org/abs/1810.05169}{{\ttfamily arXiv:1810.05169 [hep-th]}}.

\bibitem{Gonzalo:2018guu}
E.~Gonzalo, L.~E. Ib\'a\~nez, and {\'A}.~M. Uranga, ``{Modular Symmetries and
  the Swampland Conjectures},''
  \href{http://dx.doi.org/10.1007/JHEP05(2019)105}{{\em JHEP} {\bfseries 05}
  (2019) 105},
\href{http://arxiv.org/abs/1812.06520}{{\ttfamily arXiv:1812.06520 [hep-th]}}.

\bibitem{Grimm_2019}
T.~W. Grimm, C.~Li, and E.~Palti, ``{Infinite Distance Networks in Field Space
  and Charge Orbits},'' \href{http://dx.doi.org/10.1007/JHEP03(2019)016}{{\em
  JHEP} {\bfseries 03} (2019) 016},
\href{http://arxiv.org/abs/1811.02571}{{\ttfamily arXiv:1811.02571 [hep-th]}}.

\bibitem{Corvilain:2018lgw}
P.~Corvilain, T.~W. Grimm, and I.~Valenzuela, ``{The Swampland Distance
  Conjecture for K{\"a}hler moduli},''
  \href{http://dx.doi.org/10.1007/JHEP08(2019)075}{{\em JHEP} {\bfseries 08}
  (2019) 075},
\href{http://arxiv.org/abs/1812.07548}{{\ttfamily arXiv:1812.07548 [hep-th]}}.

\bibitem{grimm2019infinite}
T.~W. Grimm and D.~Van De~Heisteeg, ``{Infinite Distances and the Axion Weak
  Gravity Conjecture},'' \href{http://dx.doi.org/10.1007/JHEP03(2020)020}{{\em
  JHEP} {\bfseries 03} (2020) 020},
  \href{http://arxiv.org/abs/1905.00901}{{\ttfamily arXiv:1905.00901
  [hep-th]}}.

\bibitem{Joshi:2019nzi}
A.~Joshi and A.~Klemm, ``{Swampland Distance Conjecture for One-Parameter
  Calabi-Yau Threefolds},''
  \href{http://dx.doi.org/10.1007/JHEP08(2019)086}{{\em JHEP} {\bfseries 08}
  (2019) 086}, \href{http://arxiv.org/abs/1903.00596}{{\ttfamily
  arXiv:1903.00596 [hep-th]}}.

\bibitem{Marchesano_2019}
F.~Marchesano and M.~Wiesner, ``{Instantons and infinite distances},''
  \href{http://dx.doi.org/10.1007/JHEP08(2019)088}{{\em JHEP} {\bfseries 08}
  (2019) 088}, \href{http://arxiv.org/abs/1904.04848}{{\ttfamily
  arXiv:1904.04848 [hep-th]}}.

\bibitem{Lee:2019tst}
S.-J. Lee, W.~Lerche, and T.~Weigand, ``{Modular Fluxes, Elliptic Genera, and
  Weak Gravity Conjectures in Four Dimensions},''
  \href{http://dx.doi.org/10.1007/JHEP08(2019)104}{{\em JHEP} {\bfseries 08}
  (2019) 104}, \href{http://arxiv.org/abs/1901.08065}{{\ttfamily
  arXiv:1901.08065 [hep-th]}}.

\bibitem{Lee:2019xtm}
S.-J. Lee, W.~Lerche, and T.~Weigand, ``{Emergent Strings, Duality and Weak
  Coupling Limits for Two-Form Fields},''
\href{http://arxiv.org/abs/1904.06344}{{\ttfamily arXiv:1904.06344 [hep-th]}}.

\bibitem{lee2019emergent}
S.-J. Lee, W.~Lerche, and T.~Weigand, ``{Emergent Strings from Infinite
  Distance Limits},''
\href{http://arxiv.org/abs/1910.01135}{{\ttfamily arXiv:1910.01135 [hep-th]}}.

\bibitem{Font:2019cxq}
A.~Font, A.~Herr{\'a}ez, and L.~E. Ib{\'a}{\~n}ez, ``{The Swampland Distance
  Conjecture and Towers of Tensionless Branes},''
  \href{http://dx.doi.org/10.1007/JHEP08(2019)044}{{\em JHEP} {\bfseries 08}
  (2019) 044},
\href{http://arxiv.org/abs/1904.05379}{{\ttfamily arXiv:1904.05379 [hep-th]}}.

\bibitem{Baume:2019sry}
F.~Baume, F.~Marchesano, and M.~Wiesner, ``{Instanton Corrections and Emergent
  Strings},'' \href{http://arxiv.org/abs/1912.02218}{{\ttfamily
  arXiv:1912.02218 [hep-th]}}.

\bibitem{Blumenhagen:2017cxt}
R.~Blumenhagen, I.~Valenzuela, and F.~Wolf, ``{The Swampland Conjecture and
  F-term Axion Monodromy Inflation},''
  \href{http://dx.doi.org/10.1007/JHEP07(2017)145}{{\em JHEP} {\bfseries 07}
  (2017) 145},
\href{http://arxiv.org/abs/1703.05776}{{\ttfamily arXiv:1703.05776 [hep-th]}}.

\bibitem{Blumenhagen_2018}
R.~Blumenhagen, D.~Klaewer, L.~Schlechter, and F.~Wolf, ``{The Refined
  Swampland Distance Conjecture in Calabi-Yau Moduli Spaces},''
  \href{http://dx.doi.org/10.1007/JHEP06(2018)052}{{\em JHEP} {\bfseries 06}
  (2018) 052},
\href{http://arxiv.org/abs/1803.04989}{{\ttfamily arXiv:1803.04989 [hep-th]}}.

\bibitem{Erkinger:2019umg}
D.~Erkinger and J.~Knapp, ``{Refined swampland distance conjecture and exotic
  hybrid Calabi-Yaus},'' \href{http://dx.doi.org/10.1007/JHEP07(2019)029}{{\em
  JHEP} {\bfseries 07} (2019) 029},
  \href{http://arxiv.org/abs/1905.05225}{{\ttfamily arXiv:1905.05225
  [hep-th]}}.

\bibitem{Cecotti:2020rjq}
S.~Cecotti, ``{Special Geometry and the Swampland},''
  \href{http://arxiv.org/abs/2004.06929}{{\ttfamily arXiv:2004.06929
  [hep-th]}}.

\bibitem{Gendler:2020dfp}
N.~Gendler and I.~Valenzuela, ``{Merging the Weak Gravity and Distance
  Conjectures Using BPS Extremal Black Holes},''
  \href{http://arxiv.org/abs/2004.10768}{{\ttfamily arXiv:2004.10768
  [hep-th]}}.

\bibitem{Lanza:2020qmt}
S.~Lanza, F.~Marchesano, L.~Martucci, and I.~Valenzuela, ``{Swampland
  Conjectures for Strings and Membranes},''
  \href{http://arxiv.org/abs/2006.15154}{{\ttfamily arXiv:2006.15154
  [hep-th]}}.

\bibitem{Klaewer:2020lfg}
D.~Klaewer, S.-J. Lee, T.~Weigand, and M.~Wiesner, ``{Quantum Corrections in 4d
  N=1 Infinite Distance Limits and the Weak Gravity Conjecture},''
  \href{http://arxiv.org/abs/2011.00024}{{\ttfamily arXiv:2011.00024
  [hep-th]}}.

\bibitem{Bedroya:2019snp}
A.~Bedroya and C.~Vafa, ``{Trans-Planckian Censorship and the Swampland},''
\href{http://arxiv.org/abs/1909.11063}{{\ttfamily arXiv:1909.11063 [hep-th]}}.

\bibitem{Andriot:2020lea}
D.~Andriot, N.~Cribiori, and D.~Erkinger, ``{The web of swampland conjectures
  and the TCC bound},'' \href{http://dx.doi.org/10.1007/JHEP07(2020)162}{{\em
  JHEP} {\bfseries 07} (2020) 162},
  \href{http://arxiv.org/abs/2004.00030}{{\ttfamily arXiv:2004.00030
  [hep-th]}}.

\bibitem{Bedroya:2020rmd}
A.~Bedroya, ``{de Sitter Complementarity, TCC, and the Swampland},''
  \href{http://arxiv.org/abs/2010.09760}{{\ttfamily arXiv:2010.09760
  [hep-th]}}.

\bibitem{Lust:2019zwm}
D.~L{\"u}st, E.~Palti, and C.~Vafa, ``{AdS and the Swampland},''
  \href{http://dx.doi.org/10.1016/j.physletb.2019.134867}{{\em Physics Letters
  B} {\bfseries 797} (2019) 134867},
\href{http://arxiv.org/abs/1906.05225}{{\ttfamily arXiv:1906.05225 [hep-th]}}.

\bibitem{ElShowk:2012ht}
S.~El-Showk, M.~F. Paulos, D.~Poland, S.~Rychkov, D.~Simmons-Duffin, and
  A.~Vichi, ``{Solving the 3D Ising Model with the Conformal Bootstrap},''
  \href{http://dx.doi.org/10.1103/PhysRevD.86.025022}{{\em Phys.Rev.}
  {\bfseries D86} (2012) 025022},
\href{http://arxiv.org/abs/1203.6064}{{\ttfamily arXiv:1203.6064 [hep-th]}}.

\bibitem{El-Showk:2014dwa}
S.~El-Showk, M.~F. Paulos, D.~Poland, S.~Rychkov, D.~Simmons-Duffin, {\em
  et~al.}, ``{Solving the 3d Ising Model with the Conformal Bootstrap II.
  c-Minimization and Precise Critical Exponents},''
  \href{http://dx.doi.org/10.1007/s10955-014-1042-7}{{\em J.Stat.Phys.}
  {\bfseries 157} (2014) 869},
\href{http://arxiv.org/abs/1403.4545}{{\ttfamily arXiv:1403.4545 [hep-th]}}.

\bibitem{Kos:2016ysd}
F.~Kos, D.~Poland, D.~Simmons-Duffin, and A.~Vichi, ``{Precision islands in the
  Ising and O(N) models},''
  \href{http://dx.doi.org/10.1007/JHEP08(2016)036}{{\em JHEP} {\bfseries 08}
  (2016) 036},
\href{http://arxiv.org/abs/1603.04436}{{\ttfamily arXiv:1603.04436 [hep-th]}}.

\bibitem{dsdi}
D.~Simmons-Duffin, ``{The Lightcone Bootstrap and the Spectrum of the 3d Ising
  CFT},'' \href{http://dx.doi.org/10.1007/JHEP03(2017)086}{{\em JHEP}
  {\bfseries 03} (2017) 086},
\href{http://arxiv.org/abs/1612.08471}{{\ttfamily arXiv:1612.08471 [hep-th]}}.

\bibitem{Caron-Huot:2020ouj}
S.~Caron-Huot, Y.~Gobeil, and Z.~Zahraee, ``{The leading trajectory in the 2+1D
  Ising CFT},'' \href{http://arxiv.org/abs/2007.11647}{{\ttfamily
  arXiv:2007.11647 [hep-th]}}.

\bibitem{Beem:2013qxa}
C.~Beem, L.~Rastelli, and B.~C. van Rees, ``{The $\mathcal{N}=4$ Superconformal
  Bootstrap},'' \href{http://dx.doi.org/10.1103/PhysRevLett.111.071601}{{\em
  Phys.Rev.Lett.} {\bfseries 111} (2013) 071601},
\href{http://arxiv.org/abs/1304.1803}{{\ttfamily arXiv:1304.1803 [hep-th]}}.

\bibitem{Beem:2016wfs}
C.~Beem, L.~Rastelli, and B.~C. van Rees, ``{More ${\mathcal N}=4$
  superconformal bootstrap},''
  \href{http://dx.doi.org/10.1103/PhysRevD.96.046014}{{\em Phys. Rev. D}
  {\bfseries 96} no.~4, (2017) 046014},
  \href{http://arxiv.org/abs/1612.02363}{{\ttfamily arXiv:1612.02363
  [hep-th]}}.

\bibitem{Lin:2015wcg}
Y.-H. Lin, S.-H. Shao, D.~Simmons-Duffin, Y.~Wang, and X.~Yin, ``{N=4
  Superconformal Bootstrap of the K3 CFT},''
\href{http://arxiv.org/abs/1511.04065}{{\ttfamily arXiv:1511.04065 [hep-th]}}.

\bibitem{Behan:2017mwi}
C.~Behan, ``{Conformal manifolds: ODEs from OPEs},''
  \href{http://dx.doi.org/10.1007/JHEP03(2018)127}{{\em JHEP} {\bfseries 03}
  (2018) 127}, \href{http://arxiv.org/abs/1709.03967}{{\ttfamily
  arXiv:1709.03967 [hep-th]}}.

\bibitem{Bashmakov:2017rko}
V.~Bashmakov, M.~Bertolini, and H.~Raj, ``{On non-supersymmetric conformal
  manifolds: field theory and holography},''
  \href{http://dx.doi.org/10.1007/JHEP11(2017)167}{{\em JHEP} {\bfseries 11}
  (2017) 167}, \href{http://arxiv.org/abs/1709.01749}{{\ttfamily
  arXiv:1709.01749 [hep-th]}}.

\bibitem{Baggio:2017mas}
M.~Baggio, N.~Bobev, S.~M. Chester, E.~Lauria, and S.~S. Pufu, ``{Decoding a
  Three-Dimensional Conformal Manifold},''
  \href{http://dx.doi.org/10.1007/JHEP02(2018)062}{{\em JHEP} {\bfseries 02}
  (2018) 062}, \href{http://arxiv.org/abs/1712.02698}{{\ttfamily
  arXiv:1712.02698 [hep-th]}}.

\bibitem{Kaidi:2020ecu}
J.~Kaidi and E.~Perlmutter, ``{Discreteness and Integrality in Conformal Field
  Theory},'' \href{http://arxiv.org/abs/2008.02190}{{\ttfamily arXiv:2008.02190
  [hep-th]}}.

\bibitem{Maldacena:2011jn}
J.~Maldacena and A.~Zhiboedov, ``{Constraining Conformal Field Theories with A
  Higher Spin Symmetry},''
  \href{http://dx.doi.org/10.1088/1751-8113/46/21/214011}{{\em J. Phys.}
  {\bfseries A46} (2013) 214011},
\href{http://arxiv.org/abs/1112.1016}{{\ttfamily arXiv:1112.1016 [hep-th]}}.

\bibitem{Kontsevich:2000yf}
M.~Kontsevich and Y.~Soibelman, ``{Homological mirror symmetry and torus
  fibrations},'' in {\em {KIAS Annual International Conference on Symplectic
  Geometry and Mirror Symmetry}}, pp.~203--263.
\newblock 11, 2000.
\newblock \href{http://arxiv.org/abs/math/0011041}{{\ttfamily
  arXiv:math/0011041}}.

\bibitem{KS}
Y.~Soibelman, ``Collapsing conformal field theories and quantum spaces with
  non-negative ricci curvature.''
\newblock \url{https://www.math.ksu.edu/~soibel/nc-riem-3.pdf}.

\bibitem{Acharya:2006zw}
B.~S. Acharya and M.~R. Douglas, ``{A Finite landscape?},''
\href{http://arxiv.org/abs/hep-th/0606212}{{\ttfamily arXiv:hep-th/0606212
  [hep-th]}}.

\bibitem{Douglas:2010ic}
M.~R. Douglas, ``{Spaces of Quantum Field Theories},''
  \href{http://dx.doi.org/10.1088/1742-6596/462/1/012011}{{\em J. Phys. Conf.
  Ser.} {\bfseries 462} no.~1, (2013) 012011},
  \href{http://arxiv.org/abs/1005.2779}{{\ttfamily arXiv:1005.2779 [hep-th]}}.

\bibitem{Baume:2020dqd}
F.~Baume and J.~Calder\'on~Infante, ``{Tackling the SDC in AdS with CFTs},''
  \href{http://arxiv.org/abs/2011.03583}{{\ttfamily arXiv:2011.03583
  [hep-th]}}.

\bibitem{Zamolodchikov:1986gt}
A.~Zamolodchikov, ``{Irreversibility of the Flux of the Renormalization Group
  in a 2D Field Theory},'' {\em JETP Lett.} {\bfseries 43} (1986) 730--732.

\bibitem{Argyres:2007cn}
P.~C. Argyres and N.~Seiberg, ``{S-duality in N=2 supersymmetric gauge
  theories},'' \href{http://dx.doi.org/10.1088/1126-6708/2007/12/088}{{\em
  JHEP} {\bfseries 12} (2007) 088},
  \href{http://arxiv.org/abs/0711.0054}{{\ttfamily arXiv:0711.0054 [hep-th]}}.

\bibitem{Gaiotto:2009we}
D.~Gaiotto, ``{N=2 dualities},''
  \href{http://dx.doi.org/10.1007/JHEP08(2012)034}{{\em JHEP} {\bfseries 08}
  (2012) 034}, \href{http://arxiv.org/abs/0904.2715}{{\ttfamily arXiv:0904.2715
  [hep-th]}}.

\bibitem{Maldacena:2012sf}
J.~Maldacena and A.~Zhiboedov, ``{Constraining conformal field theories with a
  slightly broken higher spin symmetry},''
  \href{http://dx.doi.org/10.1088/0264-9381/30/10/104003}{{\em
  Class.Quant.Grav.} {\bfseries 30} (2013) 104003},
\href{http://arxiv.org/abs/1204.3882}{{\ttfamily arXiv:1204.3882 [hep-th]}}.

\bibitem{Stanev:2013qra}
Y.~S. Stanev, ``{Constraining conformal field theory with higher spin symmetry
  in four dimensions},''
  \href{http://dx.doi.org/10.1016/j.nuclphysb.2013.09.002}{{\em Nucl. Phys. B}
  {\bfseries 876} (2013) 651--666},
  \href{http://arxiv.org/abs/1307.5209}{{\ttfamily arXiv:1307.5209 [hep-th]}}.

\bibitem{Boulanger:2013zza}
N.~Boulanger, D.~Ponomarev, E.~Skvortsov, and M.~Taronna, ``{On the uniqueness
  of higher-spin symmetries in AdS and CFT},''
  \href{http://dx.doi.org/10.1142/S0217751X13501625}{{\em Int. J. Mod. Phys. A}
  {\bfseries 28} (2013) 1350162},
  \href{http://arxiv.org/abs/1305.5180}{{\ttfamily arXiv:1305.5180 [hep-th]}}.

\bibitem{Hartman:2015lfa}
T.~Hartman, S.~Jain, and S.~Kundu, ``{Causality Constraints in Conformal Field
  Theory},'' \href{http://dx.doi.org/10.1007/JHEP05(2016)099}{{\em JHEP}
  {\bfseries 05} (2016) 099},
\href{http://arxiv.org/abs/1509.00014}{{\ttfamily arXiv:1509.00014 [hep-th]}}.

\bibitem{Li:2015itl}
D.~Li, D.~Meltzer, and D.~Poland, ``{Conformal Collider Physics from the
  Lightcone Bootstrap},'' \href{http://dx.doi.org/10.1007/JHEP02(2016)143}{{\em
  JHEP} {\bfseries 02} (2016) 143},
\href{http://arxiv.org/abs/1511.08025}{{\ttfamily arXiv:1511.08025 [hep-th]}}.

\bibitem{Alba:2015upa}
V.~Alba and K.~Diab, ``{Constraining conformal field theories with a higher
  spin symmetry in $d> 3$ dimensions},''
\href{http://arxiv.org/abs/1510.02535}{{\ttfamily arXiv:1510.02535 [hep-th]}}.

\bibitem{Meltzer:2018tnm}
D.~Meltzer, ``{Higher Spin ANEC and the Space of CFTs},''
  \href{http://dx.doi.org/10.1007/JHEP07(2019)001}{{\em JHEP} {\bfseries 07}
  (2019) 001},
\href{http://arxiv.org/abs/1811.01913}{{\ttfamily arXiv:1811.01913 [hep-th]}}.

\bibitem{Roggenkamp:2003qp}
D.~Roggenkamp and K.~Wendland, ``{Limits and degenerations of unitary conformal
  field theories},'' \href{http://dx.doi.org/10.1007/s00220-004-1131-6}{{\em
  Commun. Math. Phys.} {\bfseries 251} (2004) 589--643},
  \href{http://arxiv.org/abs/hep-th/0308143}{{\ttfamily arXiv:hep-th/0308143}}.

\bibitem{Roggenkamp:2008jm}
D.~Roggenkamp and K.~Wendland, ``{Decoding the geometry of conformal field
  theories},'' {\em Bulg. J. Phys.} {\bfseries 35} (2008) 139--150,
  \href{http://arxiv.org/abs/0803.0657}{{\ttfamily arXiv:0803.0657 [hep-th]}}.

\bibitem{Papadodimas:2009eu}
K.~Papadodimas, ``{Topological Anti-Topological Fusion in Four-Dimensional
  Superconformal Field Theories},''
  \href{http://dx.doi.org/10.1007/JHEP08(2010)118}{{\em JHEP} {\bfseries 08}
  (2010) 118}, \href{http://arxiv.org/abs/0910.4963}{{\ttfamily arXiv:0910.4963
  [hep-th]}}.

\bibitem{Anselmi:1998ms}
D.~Anselmi, ``{The N=4 quantum conformal algebra},''
  \href{http://dx.doi.org/10.1016/S0550-3213(98)00848-7}{{\em Nucl. Phys. B}
  {\bfseries 541} (1999) 369--385},
  \href{http://arxiv.org/abs/hep-th/9809192}{{\ttfamily arXiv:hep-th/9809192}}.

\bibitem{Kotikov:2001sc}
A.~Kotikov and L.~Lipatov, ``{DGLAP and BFKL evolution equations in the N=4
  supersymmetric gauge theory},'' in {\em {35th Annual Winter School on Nuclear
  and Particle Physics}}.
\newblock 12, 2001.
\newblock \href{http://arxiv.org/abs/hep-ph/0112346}{{\ttfamily
  arXiv:hep-ph/0112346}}.

\bibitem{Dolan:2001tt}
F.~Dolan and H.~Osborn, ``{Superconformal symmetry, correlation functions and
  the operator product expansion},''
  \href{http://dx.doi.org/10.1016/S0550-3213(02)00096-2}{{\em Nucl.Phys.}
  {\bfseries B629} (2002) 3--73},
\href{http://arxiv.org/abs/hep-th/0112251}{{\ttfamily arXiv:hep-th/0112251
  [hep-th]}}.

\bibitem{Caron-Huot:2017vep}
S.~Caron-Huot, ``{Analyticity in Spin in Conformal Theories},''
\href{http://arxiv.org/abs/1703.00278}{{\ttfamily arXiv:1703.00278 [hep-th]}}.

\bibitem{Costa:2017twz}
M.~S. Costa, T.~Hansen, and J.~Penedones, ``{Bounds for OPE coefficients on the
  Regge trajectory},''
\href{http://arxiv.org/abs/1707.07689}{{\ttfamily arXiv:1707.07689 [hep-th]}}.

\bibitem{Behan:2017emf}
C.~Behan, L.~Rastelli, S.~Rychkov, and B.~Zan, ``{A scaling theory for the
  long-range to short-range crossover and an infrared duality},''
  \href{http://dx.doi.org/10.1088/1751-8121/aa8099}{{\em J. Phys. A} {\bfseries
  50} no.~35, (2017) 354002}, \href{http://arxiv.org/abs/1703.05325}{{\ttfamily
  arXiv:1703.05325 [hep-th]}}.

\bibitem{DiPietro:2019hqe}
L.~Di~Pietro, D.~Gaiotto, E.~Lauria, and J.~Wu, ``{3d Abelian Gauge Theories at
  the Boundary},'' \href{http://dx.doi.org/10.1007/JHEP05(2019)091}{{\em JHEP}
  {\bfseries 05} (2019) 091}, \href{http://arxiv.org/abs/1902.09567}{{\ttfamily
  arXiv:1902.09567 [hep-th]}}.

\bibitem{Schwarz:2004yj}
J.~H. Schwarz, ``{Superconformal Chern-Simons theories},''
  \href{http://dx.doi.org/10.1088/1126-6708/2004/11/078}{{\em JHEP} {\bfseries
  11} (2004) 078}, \href{http://arxiv.org/abs/hep-th/0411077}{{\ttfamily
  arXiv:hep-th/0411077}}.

\bibitem{Gaiotto:2007qi}
D.~Gaiotto and X.~Yin, ``{Notes on superconformal Chern-Simons-Matter
  theories},'' \href{http://dx.doi.org/10.1088/1126-6708/2007/08/056}{{\em
  JHEP} {\bfseries 08} (2007) 056},
  \href{http://arxiv.org/abs/0704.3740}{{\ttfamily arXiv:0704.3740 [hep-th]}}.

\bibitem{Aharony:2008ug}
O.~Aharony, O.~Bergman, D.~L. Jafferis, and J.~Maldacena, ``{N=6 superconformal
  Chern-Simons-matter theories, M2-branes and their gravity duals},''
  \href{http://dx.doi.org/10.1088/1126-6708/2008/10/091}{{\em JHEP} {\bfseries
  10} (2008) 091},
\href{http://arxiv.org/abs/0806.1218}{{\ttfamily arXiv:0806.1218 [hep-th]}}.

\bibitem{Aharony:2008gk}
O.~Aharony, O.~Bergman, and D.~L. Jafferis, ``{Fractional M2-branes},''
  \href{http://dx.doi.org/10.1088/1126-6708/2008/11/043}{{\em JHEP} {\bfseries
  11} (2008) 043},
\href{http://arxiv.org/abs/0807.4924}{{\ttfamily arXiv:0807.4924 [hep-th]}}.

\bibitem{Chang:2010sg}
C.-M. Chang and X.~Yin, ``{Families of Conformal Fixed Points of N=2
  Chern-Simons-Matter Theories},''
  \href{http://dx.doi.org/10.1007/JHEP05(2010)108}{{\em JHEP} {\bfseries 05}
  (2010) 108}, \href{http://arxiv.org/abs/1002.0568}{{\ttfamily arXiv:1002.0568
  [hep-th]}}.

\bibitem{Giombi:2011kc}
S.~Giombi, S.~Minwalla, S.~Prakash, S.~P. Trivedi, S.~R. Wadia, and X.~Yin,
  ``{Chern-Simons Theory with Vector Fermion Matter},''
  \href{http://dx.doi.org/10.1140/epjc/s10052-012-2112-0}{{\em Eur. Phys. J.}
  {\bfseries C72} (2012) 2112},
\href{http://arxiv.org/abs/1110.4386}{{\ttfamily arXiv:1110.4386 [hep-th]}}.

\bibitem{Aharony:2011jz}
O.~Aharony, G.~Gur-Ari, and R.~Yacoby, ``{d=3 Bosonic Vector Models Coupled to
  Chern-Simons Gauge Theories},''
  \href{http://dx.doi.org/10.1007/JHEP03(2012)037}{{\em JHEP} {\bfseries 03}
  (2012) 037}, \href{http://arxiv.org/abs/1110.4382}{{\ttfamily arXiv:1110.4382
  [hep-th]}}.

\bibitem{Chang:2012kt}
C.-M. Chang, S.~Minwalla, T.~Sharma, and X.~Yin, ``{ABJ Triality: from Higher
  Spin Fields to Strings},''
  \href{http://dx.doi.org/10.1088/1751-8113/46/21/214009}{{\em J. Phys. A}
  {\bfseries 46} (2013) 214009},
  \href{http://arxiv.org/abs/1207.4485}{{\ttfamily arXiv:1207.4485 [hep-th]}}.

\bibitem{Cordova:2016emh}
C.~Cordova, T.~T. Dumitrescu, and K.~Intriligator, ``{Multiplets of
  Superconformal Symmetry in Diverse Dimensions},''
  \href{http://dx.doi.org/10.1007/JHEP03(2019)163}{{\em JHEP} {\bfseries 03}
  (2019) 163}, \href{http://arxiv.org/abs/1612.00809}{{\ttfamily
  arXiv:1612.00809 [hep-th]}}.

\bibitem{Cordova:2016xhm}
C.~Cordova, T.~T. Dumitrescu, and K.~Intriligator, ``{Deformations of
  Superconformal Theories},''
  \href{http://dx.doi.org/10.1007/JHEP11(2016)135}{{\em JHEP} {\bfseries 11}
  (2016) 135}, \href{http://arxiv.org/abs/1602.01217}{{\ttfamily
  arXiv:1602.01217 [hep-th]}}.

\bibitem{Louis:2015mka}
J.~Louis and S.~L\"ust, ``{Supersymmetric AdS$_{7}$ backgrounds in half-maximal
  supergravity and marginal operators of (1, 0) SCFTs},''
  \href{http://dx.doi.org/10.1007/JHEP10(2015)120}{{\em JHEP} {\bfseries 10}
  (2015) 120}, \href{http://arxiv.org/abs/1506.08040}{{\ttfamily
  arXiv:1506.08040 [hep-th]}}.

\bibitem{Nahm:1977tg}
W.~Nahm, ``{Supersymmetries and their Representations},''
  \href{http://dx.doi.org/10.1016/0550-3213(78)90218-3}{{\em Nucl. Phys. B}
  {\bfseries 135} (1978) 149}.

\bibitem{Asnin:2009xx}
V.~Asnin, ``{On metric geometry of conformal moduli spaces of four-dimensional
  superconformal theories},''
  \href{http://dx.doi.org/10.1007/JHEP09(2010)012}{{\em JHEP} {\bfseries 09}
  (2010) 012}, \href{http://arxiv.org/abs/0912.2529}{{\ttfamily arXiv:0912.2529
  [hep-th]}}.

\bibitem{Dolan:2002zh}
F.~Dolan and H.~Osborn, ``{On short and semi-short representations for
  four-dimensional superconformal symmetry},''
  \href{http://dx.doi.org/10.1016/S0003-4916(03)00074-5}{{\em Annals Phys.}
  {\bfseries 307} (2003) 41--89},
  \href{http://arxiv.org/abs/hep-th/0209056}{{\ttfamily arXiv:hep-th/0209056}}.

\bibitem{Baggio:2014ioa}
M.~Baggio, V.~Niarchos, and K.~Papadodimas, ``{tt$^{*}$ equations, localization
  and exact chiral rings in 4d $ \mathcal{N} $ =2 SCFTs},''
  \href{http://dx.doi.org/10.1007/JHEP02(2015)122}{{\em JHEP} {\bfseries 02}
  (2015) 122}, \href{http://arxiv.org/abs/1409.4212}{{\ttfamily arXiv:1409.4212
  [hep-th]}}.

\bibitem{Gerchkovitz:2014gta}
E.~Gerchkovitz, J.~Gomis, and Z.~Komargodski, ``{Sphere Partition Functions and
  the Zamolodchikov Metric},''
  \href{http://dx.doi.org/10.1007/JHEP11(2014)001}{{\em JHEP} {\bfseries 11}
  (2014) 001}, \href{http://arxiv.org/abs/1405.7271}{{\ttfamily arXiv:1405.7271
  [hep-th]}}.

\bibitem{Gomis:2015yaa}
J.~Gomis, P.-S. Hsin, Z.~Komargodski, A.~Schwimmer, N.~Seiberg, and S.~Theisen,
  ``{Anomalies, Conformal Manifolds, and Spheres},''
  \href{http://dx.doi.org/10.1007/JHEP03(2016)022}{{\em JHEP} {\bfseries 03}
  (2016) 022}, \href{http://arxiv.org/abs/1509.08511}{{\ttfamily
  arXiv:1509.08511 [hep-th]}}.

\bibitem{Tachikawa:2017aux}
Y.~Tachikawa and K.~Yonekura, ``{Anomalies involving the space of couplings and
  the Zamolodchikov metric},''
  \href{http://dx.doi.org/10.1007/JHEP12(2017)140}{{\em JHEP} {\bfseries 12}
  (017) 140}, \href{http://arxiv.org/abs/1710.03934}{{\ttfamily
  arXiv:1710.03934 [hep-th]}}.

\bibitem{Seiberg:2018ntt}
N.~Seiberg, Y.~Tachikawa, and K.~Yonekura, ``{Anomalies of Duality Groups and
  Extended Conformal Manifolds},''
  \href{http://dx.doi.org/10.1093/ptep/pty069}{{\em PTEP} {\bfseries 2018}
  no.~7, (2018) 073B04}, \href{http://arxiv.org/abs/1803.07366}{{\ttfamily
  arXiv:1803.07366 [hep-th]}}.

\bibitem{Beem:2014zpa}
C.~Beem, M.~Lemos, P.~Liendo, L.~Rastelli, and B.~C. van Rees, ``{The
  ${\mathcal N}=2$ superconformal bootstrap},''
\href{http://arxiv.org/abs/1412.7541}{{\ttfamily arXiv:1412.7541 [hep-th]}}.

\bibitem{Pestun:2007rz}
V.~Pestun, ``{Localization of gauge theory on a four-sphere and supersymmetric
  Wilson loops},'' \href{http://dx.doi.org/10.1007/s00220-012-1485-0}{{\em
  Commun. Math. Phys.} {\bfseries 313} (2012) 71--129},
  \href{http://arxiv.org/abs/0712.2824}{{\ttfamily arXiv:0712.2824 [hep-th]}}.

\bibitem{Baggio:2015vxa}
M.~Baggio, V.~Niarchos, and K.~Papadodimas, ``{On exact correlation functions
  in SU(N) $ \mathcal{N}=2 $ superconformal QCD},''
  \href{http://dx.doi.org/10.1007/JHEP11(2015)198}{{\em JHEP} {\bfseries 11}
  (2015) 198}, \href{http://arxiv.org/abs/1508.03077}{{\ttfamily
  arXiv:1508.03077 [hep-th]}}.

\bibitem{Gerchkovitz:2016gxx}
E.~Gerchkovitz, J.~Gomis, N.~Ishtiaque, A.~Karasik, Z.~Komargodski, and S.~S.
  Pufu, ``{Correlation Functions of Coulomb Branch Operators},''
  \href{http://dx.doi.org/10.1007/JHEP01(2017)103}{{\em JHEP} {\bfseries 01}
  (2017) 103}, \href{http://arxiv.org/abs/1602.05971}{{\ttfamily
  arXiv:1602.05971 [hep-th]}}.

\bibitem{Grassi:2019txd}
A.~Grassi, Z.~Komargodski, and L.~Tizzano, ``{Extremal Correlators and Random
  Matrix Theory},'' \href{http://arxiv.org/abs/1908.10306}{{\ttfamily
  arXiv:1908.10306 [hep-th]}}.

\bibitem{Kastor:1988ef}
D.~Kastor, E.~Martinec, and S.~Shenker, ``{RG Flow in N=1 Discrete Series},''
  \href{http://dx.doi.org/10.1016/0550-3213(89)90060-6}{{\em Nucl. Phys. B}
  {\bfseries 316} (1989) 590--608}.

\bibitem{Meade:2008wd}
P.~Meade, N.~Seiberg, and D.~Shih, ``{General Gauge Mediation},''
  \href{http://dx.doi.org/10.1143/PTPS.177.143}{{\em Prog. Theor. Phys. Suppl.}
  {\bfseries 177} (2009) 143--158},
  \href{http://arxiv.org/abs/0801.3278}{{\ttfamily arXiv:0801.3278 [hep-ph]}}.

\bibitem{Leigh:1995ep}
R.~G. Leigh and M.~J. Strassler, ``{Exactly marginal operators and duality in
  four-dimensional N=1 supersymmetric gauge theory},''
  \href{http://dx.doi.org/10.1016/0550-3213(95)00261-P}{{\em Nucl. Phys. B}
  {\bfseries 447} (1995) 95--136},
  \href{http://arxiv.org/abs/hep-th/9503121}{{\ttfamily arXiv:hep-th/9503121}}.

\bibitem{Green:2010da}
D.~Green, Z.~Komargodski, N.~Seiberg, Y.~Tachikawa, and B.~Wecht, ``{Exactly
  Marginal Deformations and Global Symmetries},''
  \href{http://dx.doi.org/10.1007/JHEP06(2010)106}{{\em JHEP} {\bfseries 06}
  (2010) 106}, \href{http://arxiv.org/abs/1005.3546}{{\ttfamily arXiv:1005.3546
  [hep-th]}}.

\bibitem{Razamat:2019vfd}
S.~S. Razamat and G.~Zafrir, ``{$N=1$ conformal dualities},''
  \href{http://dx.doi.org/10.1007/JHEP09(2019)046}{{\em JHEP} {\bfseries 09}
  (2019) 046}, \href{http://arxiv.org/abs/1906.05088}{{\ttfamily
  arXiv:1906.05088 [hep-th]}}.

\bibitem{Razamat:2020gcc}
S.~S. Razamat and G.~Zafrir, ``{$ \mathcal{N} $ = 1 conformal duals of gauged
  E$_{n}$ MN models},'' \href{http://dx.doi.org/10.1007/JHEP06(2020)176}{{\em
  JHEP} {\bfseries 06} (2020) 176},
  \href{http://arxiv.org/abs/2003.01843}{{\ttfamily arXiv:2003.01843
  [hep-th]}}.

\bibitem{Razamat:2020pra}
S.~S. Razamat, E.~Sabag, and G.~Zafrir, ``{Weakly coupled conformal manifolds
  in 4d},'' \href{http://dx.doi.org/10.1007/JHEP06(2020)179}{{\em JHEP}
  {\bfseries 06} (2020) 179}, \href{http://arxiv.org/abs/2004.07097}{{\ttfamily
  arXiv:2004.07097 [hep-th]}}.

\bibitem{Komargodski:2020ved}
Z.~Komargodski, S.~S. Razamat, O.~Sela, and A.~Sharon, ``{A Nilpotency Index of
  Conformal Manifolds},'' \href{http://dx.doi.org/10.1007/JHEP10(2020)183}{{\em
  JHEP} {\bfseries 10} (2020) 183},
  \href{http://arxiv.org/abs/2003.04579}{{\ttfamily arXiv:2003.04579
  [hep-th]}}.

\bibitem{Aharony:2002hx}
O.~Aharony, B.~Kol, and S.~Yankielowicz, ``{On exactly marginal deformations of
  N=4 SYM and type IIB supergravity on AdS(5) x S**5},''
  \href{http://dx.doi.org/10.1088/1126-6708/2002/06/039}{{\em JHEP} {\bfseries
  06} (2002) 039}, \href{http://arxiv.org/abs/hep-th/0205090}{{\ttfamily
  arXiv:hep-th/0205090}}.

\bibitem{Buican:2014sfa}
M.~Buican and T.~Nishinaka, ``{Compact Conformal Manifolds},''
  \href{http://dx.doi.org/10.1007/JHEP01(2015)112}{{\em JHEP} {\bfseries 01}
  (2015) 112}, \href{http://arxiv.org/abs/1410.3006}{{\ttfamily arXiv:1410.3006
  [hep-th]}}.

\bibitem{Bachas:2019jaa}
C.~Bachas, I.~Lavdas, and B.~Le~Floch, ``{Marginal Deformations of 3d $N=4$
  Linear Quiver Theories},''
  \href{http://dx.doi.org/10.1007/JHEP10(2019)253}{{\em JHEP} {\bfseries 10}
  (2019) 253}, \href{http://arxiv.org/abs/1905.06297}{{\ttfamily
  arXiv:1905.06297 [hep-th]}}.

\bibitem{Beratto:2020qyk}
E.~Beratto, N.~Mekareeya, and M.~Sacchi, ``{Marginal operators and
  supersymmetry enhancement in 3d $S$-fold SCFTs},''
  \href{http://arxiv.org/abs/2009.10123}{{\ttfamily arXiv:2009.10123
  [hep-th]}}.

\bibitem{Vasiliev:1999ba}
M.~A. Vasiliev, ``{Higher spin gauge theories: Star product and AdS space},''
\href{http://arxiv.org/abs/hep-th/9910096}{{\ttfamily arXiv:hep-th/9910096
  [hep-th]}}.

\bibitem{Klebanov:2002ja}
I.~R. Klebanov and A.~M. Polyakov, ``{AdS dual of the critical O(N) vector
  model},'' \href{http://dx.doi.org/10.1016/S0370-2693(02)02980-5}{{\em Phys.
  Lett.} {\bfseries B550} (2002) 213--219},
\href{http://arxiv.org/abs/hep-th/0210114}{{\ttfamily arXiv:hep-th/0210114
  [hep-th]}}.

\bibitem{Gaberdiel:2010pz}
M.~R. Gaberdiel and R.~Gopakumar, ``{An AdS$_{3}$ Dual for Minimal Model
  CFTs},'' \href{http://dx.doi.org/10.1103/PhysRevD.83.066007}{{\em Phys. Rev.
  D} {\bfseries 83} (2011) 066007},
  \href{http://arxiv.org/abs/1011.2986}{{\ttfamily arXiv:1011.2986 [hep-th]}}.

\bibitem{Shenker:2011zf}
S.~H. Shenker and X.~Yin, ``{Vector Models in the Singlet Sector at Finite
  Temperature},'' \href{http://arxiv.org/abs/1109.3519}{{\ttfamily
  arXiv:1109.3519 [hep-th]}}.

\bibitem{Giombi:2012ms}
S.~Giombi and X.~Yin, ``{The Higher Spin/Vector Model Duality},''
  \href{http://dx.doi.org/10.1088/1751-8113/46/21/214003}{{\em J. Phys. A}
  {\bfseries 46} (2013) 214003},
  \href{http://arxiv.org/abs/1208.4036}{{\ttfamily arXiv:1208.4036 [hep-th]}}.

\bibitem{Didenko:2012tv}
V.~Didenko and E.~Skvortsov, ``{Exact higher-spin symmetry in CFT: all
  correlators in unbroken Vasiliev theory},''
  \href{http://dx.doi.org/10.1007/JHEP04(2013)158}{{\em JHEP} {\bfseries 04}
  (2013) 158}, \href{http://arxiv.org/abs/1210.7963}{{\ttfamily arXiv:1210.7963
  [hep-th]}}.

\bibitem{Sundborg:2000wp}
B.~Sundborg, ``{Stringy gravity, interacting tensionless strings and massless
  higher spins},'' \href{http://dx.doi.org/10.1016/S0920-5632(01)01545-6}{{\em
  Nucl. Phys. B Proc. Suppl.} {\bfseries 102} (2001) 113--119},
  \href{http://arxiv.org/abs/hep-th/0103247}{{\ttfamily arXiv:hep-th/0103247}}.

\bibitem{HaggiMani:2000ru}
P.~Haggi-Mani and B.~Sundborg, ``{Free large N supersymmetric Yang-Mills theory
  as a string theory},''
  \href{http://dx.doi.org/10.1088/1126-6708/2000/04/031}{{\em JHEP} {\bfseries
  04} (2000) 031}, \href{http://arxiv.org/abs/hep-th/0002189}{{\ttfamily
  arXiv:hep-th/0002189}}.

\bibitem{Sezgin:2002rt}
E.~Sezgin and P.~Sundell, ``{Massless higher spins and holography},''
  \href{http://dx.doi.org/10.1016/S0550-3213(02)00739-3,
  10.1016/S0550-3213(03)00267-0}{{\em Nucl. Phys.} {\bfseries B644} (2002)
  303--370}, \href{http://arxiv.org/abs/hep-th/0205131}{{\ttfamily
  arXiv:hep-th/0205131 [hep-th]}}.
[Erratum: Nucl. Phys.B660,403(2003)].

\bibitem{Bianchi:2003wx}
M.~Bianchi, J.~F. Morales, and H.~Samtleben, ``{On stringy AdS(5) x S**5 and
  higher spin holography},''
  \href{http://dx.doi.org/10.1088/1126-6708/2003/07/062}{{\em JHEP} {\bfseries
  07} (2003) 062}, \href{http://arxiv.org/abs/hep-th/0305052}{{\ttfamily
  arXiv:hep-th/0305052}}.

\bibitem{Beisert:2003te}
N.~Beisert, M.~Bianchi, J.~Morales, and H.~Samtleben, ``{On the spectrum of AdS
  / CFT beyond supergravity},''
  \href{http://dx.doi.org/10.1088/1126-6708/2004/02/001}{{\em JHEP} {\bfseries
  02} (2004) 001}, \href{http://arxiv.org/abs/hep-th/0310292}{{\ttfamily
  arXiv:hep-th/0310292}}.

\bibitem{Gaberdiel:2015wpo}
M.~R. Gaberdiel and R.~Gopakumar, ``{String Theory as a Higher Spin Theory},''
  \href{http://dx.doi.org/10.1007/JHEP09(2016)085}{{\em JHEP} {\bfseries 09}
  (2016) 085}, \href{http://arxiv.org/abs/1512.07237}{{\ttfamily
  arXiv:1512.07237 [hep-th]}}.

\bibitem{Gaberdiel:2018rqv}
M.~R. Gaberdiel and R.~Gopakumar, ``{Tensionless string spectra on
  AdS$_{3}$},'' \href{http://dx.doi.org/10.1007/JHEP05(2018)085}{{\em JHEP}
  {\bfseries 05} (2018) 085}, \href{http://arxiv.org/abs/1803.04423}{{\ttfamily
  arXiv:1803.04423 [hep-th]}}.

\bibitem{Eberhardt:2018ouy}
L.~Eberhardt, M.~R. Gaberdiel, and R.~Gopakumar, ``{The Worldsheet Dual of the
  Symmetric Product CFT},''
  \href{http://dx.doi.org/10.1007/JHEP04(2019)103}{{\em JHEP} {\bfseries 04}
  (2019) 103}, \href{http://arxiv.org/abs/1812.01007}{{\ttfamily
  arXiv:1812.01007 [hep-th]}}.

\bibitem{Caron-Huot:2016icg}
S.~Caron-Huot, Z.~Komargodski, A.~Sever, and A.~Zhiboedov, ``{Strings from
  Massive Higher Spins: The Asymptotic Uniqueness of the Veneziano
  Amplitude},''
\href{http://arxiv.org/abs/1607.04253}{{\ttfamily arXiv:1607.04253 [hep-th]}}.

\bibitem{Freedman:1998tz}
D.~Z. Freedman, S.~D. Mathur, A.~Matusis, and L.~Rastelli, ``{Correlation
  functions in the CFT(d) / AdS(d+1) correspondence},''
  \href{http://dx.doi.org/10.1016/S0550-3213(99)00053-X}{{\em Nucl. Phys.}
  {\bfseries B546} (1999) 96--118},
\href{http://arxiv.org/abs/hep-th/9804058}{{\ttfamily arXiv:hep-th/9804058
  [hep-th]}}.

\bibitem{Myers:2010tj}
R.~C. Myers and A.~Sinha, ``{Holographic c-theorems in arbitrary dimensions},''
  \href{http://dx.doi.org/10.1007/JHEP01(2011)125}{{\em JHEP} {\bfseries 01}
  (2011) 125}, \href{http://arxiv.org/abs/1011.5819}{{\ttfamily arXiv:1011.5819
  [hep-th]}}.

\bibitem{Bhardwaj:2013qia}
L.~Bhardwaj and Y.~Tachikawa, ``{Classification of 4d N=2 gauge theories},''
  \href{http://dx.doi.org/10.1007/JHEP12(2013)100}{{\em JHEP} {\bfseries 12}
  (2013) 100}, \href{http://arxiv.org/abs/1309.5160}{{\ttfamily arXiv:1309.5160
  [hep-th]}}.

\bibitem{Gubser:1998vd}
S.~S. Gubser, ``{Einstein manifolds and conformal field theories},''
  \href{http://dx.doi.org/10.1103/PhysRevD.59.025006}{{\em Phys. Rev.}
  {\bfseries D59} (1999) 025006},
\href{http://arxiv.org/abs/hep-th/9807164}{{\ttfamily arXiv:hep-th/9807164
  [hep-th]}}.

\bibitem{bishop1964geometry}
R.~Bishop and R.~J. Crittenden, {\em Geometry of Manifolds}.
\newblock Pure and applied mathematics. Academic Press, 1964.
\newblock \url{https://books.google.com/books?id=N4EpAQAAMAAJ}.

\bibitem{Cvetic2005}
M.~Cveti\u{c}, H.~L\''{u}, D.~N. Page, and C.~N. Pope, ``New einstein-sasaki
  spaces in five and higher dimensions,''
  \href{http://dx.doi.org/10.1103/physrevlett.95.071101}{{\em Physical Review
  Letters} {\bfseries 95} no.~7, (Aug, 2005) }.
  \url{http://dx.doi.org/10.1103/PhysRevLett.95.071101}.

\bibitem{Beem:2014rza}
C.~Beem, W.~Peelaers, L.~Rastelli, and B.~C. van Rees, ``{Chiral algebras of
  class S},'' \href{http://dx.doi.org/10.1007/JHEP05(2015)020}{{\em JHEP}
  {\bfseries 05} (2015) 020},
\href{http://arxiv.org/abs/1408.6522}{{\ttfamily arXiv:1408.6522 [hep-th]}}.

\bibitem{Polchinski:1999ry}
J.~Polchinski, ``{S matrices from AdS space-time},''
  \href{http://arxiv.org/abs/hep-th/9901076}{{\ttfamily arXiv:hep-th/9901076}}.

\bibitem{Grimm:2020cda}
T.~W. Grimm, ``{Moduli Space Holography and the Finiteness of Flux Vacua},''
  \href{http://arxiv.org/abs/2010.15838}{{\ttfamily arXiv:2010.15838
  [hep-th]}}.

\bibitem{Sundborg:1999ue}
B.~Sundborg, ``{The Hagedorn transition, deconfinement and N=4 SYM theory},''
  \href{http://dx.doi.org/10.1016/S0550-3213(00)00044-4}{{\em Nucl. Phys. B}
  {\bfseries 573} (2000) 349--363},
  \href{http://arxiv.org/abs/hep-th/9908001}{{\ttfamily arXiv:hep-th/9908001}}.

\bibitem{Aharony:2003sx}
O.~Aharony, J.~Marsano, S.~Minwalla, K.~Papadodimas, and M.~Van~Raamsdonk,
  ``{The Hagedorn - deconfinement phase transition in weakly coupled large N
  gauge theories},'' \href{http://dx.doi.org/10.4310/ATMP.2004.v8.n4.a1}{{\em
  Adv. Theor. Math. Phys.} {\bfseries 8} (2004) 603--696},
  \href{http://arxiv.org/abs/hep-th/0310285}{{\ttfamily arXiv:hep-th/0310285}}.

\bibitem{Beisert:2004di}
N.~Beisert, M.~Bianchi, J.~F. Morales, and H.~Samtleben, ``{Higher spin
  symmetry and N=4 SYM},''
  \href{http://dx.doi.org/10.1088/1126-6708/2004/07/058}{{\em JHEP} {\bfseries
  07} (2004) 058}, \href{http://arxiv.org/abs/hep-th/0405057}{{\ttfamily
  arXiv:hep-th/0405057}}.

\bibitem{DHoker:2002nbb}
E.~D'Hoker and D.~Z. Freedman, ``{Supersymmetric gauge theories and the AdS /
  CFT correspondence},'' in {\em {Theoretical Advanced Study Institute in
  Elementary Particle Physics (TASI 2001): Strings, Branes and EXTRA
  Dimensions}}, pp.~3--158.
\newblock 1, 2002.
\newblock \href{http://arxiv.org/abs/hep-th/0201253}{{\ttfamily
  arXiv:hep-th/0201253}}.

\bibitem{Gadde:2009dj}
A.~Gadde, E.~Pomoni, and L.~Rastelli, ``{The Veneziano Limit of N = 2
  Superconformal QCD: Towards the String Dual of N = 2 SU(N(c)) SYM with N(f) =
  2 N(c)},'' \href{http://arxiv.org/abs/0912.4918}{{\ttfamily arXiv:0912.4918
  [hep-th]}}.

\bibitem{Lunin:2005jy}
O.~Lunin and J.~M. Maldacena, ``{Deforming field theories with U(1) x U(1)
  global symmetry and their gravity duals},''
  \href{http://dx.doi.org/10.1088/1126-6708/2005/05/033}{{\em JHEP} {\bfseries
  05} (2005) 033}, \href{http://arxiv.org/abs/hep-th/0502086}{{\ttfamily
  arXiv:hep-th/0502086}}.

\end{thebibliography}\endgroup

\end{document}